\documentclass[10pt,journal,compsoc]{IEEEtran}
\usepackage{url}
\usepackage{multirow}
\usepackage{multicol}
\usepackage{diagbox}
\ifCLASSOPTIONcompsoc
  \usepackage[nocompress]{cite}
\else
  \usepackage{cite}
\fi
\ifCLASSINFOpdf
  \usepackage{graphicx, subcaption}
\else
\fi
\usepackage{amsmath}
\DeclareMathOperator*{\argmax}{argmax} 
\DeclareMathOperator*{\argmin}{argmin}
\usepackage{amssymb}
\usepackage{amsthm}
\usepackage{algorithm}
\usepackage{algorithmic}
\graphicspath{ {./manuscript/figures/} }

\begin{document}
%
\title{Privacy-Preserving Boosting in the Local Setting}
%
%
%
%

\author{
        Sen Wang, and J. Morris Chang, Senior Member, IEEE
\thanks{}}

\IEEEtitleabstractindextext{%
\begin{abstract}
In machine learning, boosting is one of the most popular methods that designed to combine multiple base learners to a superior one. The well-known Boosted Decision Tree classifier, has been widely adopted in many areas. In the big data era, the data held by individual and entities, like personal images, browsing history and census information, are more likely to contain sensitive information. The privacy concern raises when such data leaves the hand of the owners and be further explored or mined. Such privacy issue demands that the machine learning algorithm should be privacy aware. Recently, Local Differential Privacy is proposed as an effective privacy protection approach, which offers a strong guarantee to the data owners, as the data is perturbed before any further usage, and the true values never leave the hands of the owners. Thus the machine learning algorithm with the private data instances is of great value and importance. In this paper, we are interested in developing the privacy-preserving boosting algorithm that a data user is allowed to build a classifier without knowing or deriving the exact value of each data samples. Our experiments demonstrate the effectiveness of the proposed boosting algorithm and the high utility of the learned classifiers.

\end{abstract}

}

\maketitle

\IEEEdisplaynontitleabstractindextext

%
\IEEEpeerreviewmaketitle

\IEEEraisesectionheading{\section{Introduction}\label{sec:introduction}}

%
%
%
%
\IEEEPARstart{I}{n}
%
%
%
%
machine learning, ensemble learning\cite{hansen1990neural} refers to the strategy of combining multiple hypotheses to form a better one. The superior performance over a single base learning algorithm often leads to a better prediction. Many works have been conducted under the area and there are three conventional methods, the bagging\cite{breiman1996bagging}, boosting\cite{schapire1990strength,freund1997decision} and stacking\cite{wolpert1992stacked}. Among all three, the boosting method focuses on training a set of base learners sequentially and assigning each base learner a weight, then the final decision is made by a weighted majority voting over all base learners. As an example, the Boosted Decision Tree (BDT) is of great popular and widely adopted in many different applications, like text mining\cite{apte1998text}, geographical classification\cite{pal2003assessment} and finance\cite{xia2017boosted}. Like other machine learning algorithm, the boosting algorithm is developed under the assumption of a centralized fashion, where all data has been collected together. However, in the big data era, the data is generated and stored more sparsely than before, which brings new challenges and difficulties for such algorithm, for instance, privacy protection is one of the most emerging demands in current society.

Data explosion results in tons of data are generated and held by the individual and entities, such as the personal image, financial records and census data. The privacy concern raises when the data leaves the hands of data owners and participates in some computations. The AOL search engine log\cite{shen2007privacy} and Netflix prize contest\cite{narayanan2008robust} attacks prove such threat and show the needs that machine learning algorithms should be privacy aware. As a promising solution, Differential Privacy \cite{Dwork2006DP} gives a definition of the privacy and how it can be protected. A mechanism is said to be differentially private if the computation result of the data is robust to any change of any individual sample. And several differentially private machine learning algorithms\cite{ji2014differential} has been developed since then. A trusted third party is originally introduced to gather data from individual owners and is responsible to process the data privately. Recently, Local Differential Privacy (LDP)\cite{duchi2013local,wang2017locally} takes the concept into a local setting and a mechanism is defined to be local differentially private if the processing makes any two samples indistinguishable. An advantage with LDP is that it allows the data owners to perturb the input by themselves and the true values never leave the hands of the data owner. Thus unlike DP, there is no need of trusted third party anymore.

The indistinguishability of any two data samples brings a strong privacy guarantee for the data owners, and relieve the fear of information leakage. Thus developing the machine learning algorithm over such ``private'' data samples is of great value and importance. To the best of our knowledge, there is little work about developing the boosting algorithm with the protection of LDP. In this paper, we are eager to fill such gap by developing a privacy-preserving boosting algorithm that satisfies LDP. More specifically, we consider such a problem: there are two types of parties, multiple data owners and a data user. Each data owner holds a set of training samples; the data user intends to fit a boosted classifier (e.g., BDT) with the samples from data owners. During the computation, the data owner perturbs their data using the mechanism that satisfies LDP and only pass such perturbed data to the data user. In the end, the data user should only learn the classifier without knowing or deriving the value of any individual sample from the classifier, thus the privacy of the data samples are protected.

In the meanwhile, the randomized perturbation mechanism brings noise into the training samples and the mechanism is said to satisfy $\epsilon$-LDP ($\epsilon >$ 0), where the smaller $\epsilon$ is, the more privacy preserved and the more noise injected. To maintain the utility of the learned classifier, we rely on the statistical information of the perturbed samples (e.g., mean estimation), in which the adopted randomized mechanism provides an asymptotically optimal error bound. To demonstrate, we compare the utility of the learned classifier in terms of prediction capacity given different existing  randomized mechanisms and show that the model learned by our algorithm effectively maintain a high utility.

Furthermore, beyond the widely adopted BDT classifier, the proposed boosting algorithm is capable to support other types of classifiers, as the original design of the boosting procedure. Thus we study other types of classifiers and analyze the strategies to support such classifiers in our boosting algorithm. In the experiment, three types of boosted classifiers are implemented and the performance of the classifiers are evaluated. 
  
Overall, the contributions of our work are three folds:

\begin{itemize}
\item We propose a privacy-preserving boosting algorithm, and implement the widely adopted Boosted Decision Tree classifier. To the best of our knowledge, there are few works investigating the privacy-preserving boosting mechanism with LDP protection and we provide a comprehensive study in this paper.
\item To maintain the utility, the learned classifier is built over the statistical information of the perturbed training samples, which results in an asymptotically optimal error bound. To demonstrate that, we compare multiple existing perturbation methods that satisfy $\epsilon-$LDP and show the superior performance of our method with both real and synthetic datasets.
\item Beyond the BDT, we also analyze how to support other classifiers with our boosting algorithm and summarize the type of data for corresponding classifiers, for instance, the Logistic Regression and Nearest Centroid Classifier. In the experiment, both boosted classifiers are also implemented and the utility of them are evaluated comprehensively.
\end{itemize}

The rest of the paper is organized as follow. The preliminary is in Section II. The problem definition and proposed solution are introduced in Section III. The experimental evaluation is given in Section IV. The related work is presented in Section V. Section VI provides the conclusion.

\begin{table}[h]
\small
\centering
\caption{Notations and Symbols}
\resizebox{\columnwidth}{!}{%
\begin{tabular}{c|l}
\hline
$(X^l, Y^l)$ & the dataset held by data owner $l$ \\ 
$(\boldsymbol{x_i}^l, y_i^l)$ & the $i$th sample and label held by data owner $l$ \\ 
$x_{i,j}$ & the value of the $i$th sample of the $j$th feature \\ 
$x_i'$ & the perturbed output of $x_i$ \\
$N^l$ & number of sample held by data owner $l$ \\
$L$ & the number of data owners \\
$K$ & number of classes in the dataset \\
$d$ & dimensionality of the dataset \\ 

$Pr[\cdot]$ & probability of an event \\
$\mathcal{L}()$ & loss function \\

$M$ & number of base classifiers in the boosting algorithm \\
$T_m^l$ & the $m$th base classifier fitted by data owner $l$ \\
$w_i$ & weights of the $i$th samples \\
$S^j$ & the cross table of the $j$th feature  \\ \hline
$\epsilon$-DP & $\epsilon$-differential privacy \\
$\epsilon$-LDP & $\epsilon$-local differential privacy \\
LR & logistic regression \\
DT & decision tree \\ 
BDT & boosted decision tree \\ 
NCC & nearest centroid classifier\\
MSE & mean squared error \\
\hline
\end{tabular}
}
\label{tab:notation}
\end{table}

\section{Preliminary}
\subsection{Differential Privacy}
Differential Privacy (DP) attracts lots of attention in the privacy research community in the last decades, which provides a measurement of the information leakage of individual sample from the computation over an underlying dataset. Originally, DP considers the setting that a trusted data curator gathers data from multiple data owners and performs a computation from the data, like learn the mean value or find the maximum/minimum value. To ensure no one can reliably infer any individual sample from the computation result, the curator adds random noise to the result such that the released one would not change if any sample of the underlying data changes. Since no single sample can significantly affect the distribution, adversaries cannot infer the information corresponding to any individual sample confidently. Formally, given two data databases $A,A^\ast$, it is said that $A,A^\ast$ are \textit{neighbors} if they differ on at most one row. The definition of a $(\epsilon,\delta)$-differential private mechanism over $A$ is defined below: 

\textit{Definition 1 }($\epsilon,\delta$)-Differential Privacy\cite{Dwork2006DP, dwork2014analyze}: A randomized mechanism $\mathcal{F}$ is $(\epsilon,\delta)$-differentially private if for every two neighboring databases $A,A^\ast$ and for any $\mathcal{O} \subseteq Range(\mathcal{F})$,
\begin{equation}
    Pr[\mathcal{F}(A)\in \mathcal{O}] \leq e^\epsilon Pr[\mathcal{F}(A^\ast)\in \mathcal{O}] + \delta
\end{equation}

where $Pr[\cdot]$ denotes the probability of an event, $Range(\mathcal{F})$ denotes the set of all possible outputs of the algorithm $\mathcal{F}$. The smaller $\epsilon,\delta$ are, the closer $Pr[\mathcal{F}(A)\in \mathcal{O}]$ and $Pr[\mathcal{F}(A^\ast)\in \mathcal{O}]$ are, and the stronger privacy protection gains. When $\delta=0$, the mechanism $\mathcal{F}$ satisfies $\epsilon$-DP, which is a stronger privacy guarantee than $(\epsilon,\delta)$-DP with $\delta>0$.

\subsection{Local Differential Privacy}
Similar to DP, a setting of \textit{local privacy} is investigated by Duchi et al. \cite{duchi2013local}, namely Local Differential Privacy (LDP), which provides the plausible deniability to the data owner. LDP shifts the perturbation from the central site to the local data owner. It considers a scenario that there is no trusted third party, and an untrustworthy data curator needs to collect data from data owners and perform certain computations. The data owners are still willing to contribute their data, but the  privacy of the data must be enforced. The formal definition is given below:

\textit{Definition 2 }$\epsilon$-Local Differential Privacy\cite{duchi2013local, wang2017locally}: A randomized mechanism $\mathcal{G}$ satisfies $\epsilon$-LDP if for any \textit{input} $v_1$ and $v_2$ and for any $\mathcal{O} \subseteq Range(\mathcal{G})$:
\begin{equation}
    Pr[\mathcal{G}(v_1) \in \mathcal{O}] \leq e^\epsilon Pr[\mathcal{G}(v_2) \in \mathcal{O}]
\end{equation}

Comparing to DP, LDP provides another solution for the data owners. Instead of submitting the true values, the data owners perturb the true data with the mechanism that satisfies $\epsilon$-LDP and then release the perturbed ones. Thus LDP gives a stronger privacy protection, as the curator only sees the perturbed data and the true values of the data never leave the data owners' hands.

\subsection{AdaBoosting}
Boosting\cite{schapire1990strength} is a widely used ensemble algorithm to reduce the bias and variance in supervised learning. Many boosting algorithms consist of constructing a classifier in an incremental fashion by adding weak classifiers to a pool, and using their weighted ``vote'' to determine the final classification. Usually, the weak classifier is also referred as a base classifier, the term ``weak'' means the classifier may have limited predication capacity, for example, the prediction result is slightly better than random guess. For consistency, we use base classifier in this paper, rather than weak classifier. AdaBoost\cite{freund1997decision} is one of the best learning methods developed in the last two decades. In the algorithm, each base classifier is weighted based on its prediction capacity. Initially, all samples share the same equal weight. After a base classifier is added, the sample weights are readjusted such that the misclassified samples gain a higher weights and samples that are classified correctly lose weights. Thus, subsequent base classifiers focus more on the samples that were previously misclassified. And it has been proven\cite{freund1997decision} that the final model can converge to a strong learner. Given a multi-class classification problem, where there are a set of training data $(\boldsymbol{x}_1, y_1),\dots,(\boldsymbol{x}_n, y_n)$, $\boldsymbol{x}_i \in \mathbb{R}^d$ and  $y_i\in \{1,2,\dots,K\}$ is the corresponding class label, the goal is to build a meta-classifier $T(\boldsymbol{\cdot})$ from the training data, which consists of a set of weighted base classifiers. The pseudo code of the SAMME \cite{hastie2009multi} algorithm is presented in Alg.~\ref{alg:samme}:
\begin{algorithm}
\caption{Stagewise Additive Modeling using a Multi-class Exponential loss function (SAMME)}
\label{alg:samme}
\begin{algorithmic}[1]
\STATE \textbf{Input}: $\{(\boldsymbol{x}_1, y_1),\dots,(\boldsymbol{x}_n, y_n)\}$.
\STATE Initialize the observation weights $w_i = 1/n, i=1,2,\dots,n$
\FOR{m=1 to M:}
\STATE Fit a classifier $T_m(\cdot)$ to the training data using weights $w_i$
\STATE Compute $err_m = \frac{\sum^{n}_{i=1}w_i\mathbb{I}(y_i\neq T_m(\boldsymbol{x}_i))}{\sum^{n}_{i=1}w_i}$
\STATE Compute $\alpha_m=log\frac{1-err_m}{err_m}+log(K-1)$
\STATE Set $w_i \leftarrow w_i\cdot exp(\alpha_m\cdot \mathbb{I}(y_i\neq T_m(\boldsymbol{x}_i))),i=1,2,\dots,n$
\STATE Re-normalize $w_i$
\ENDFOR

\RETURN $\{(\alpha_1,T_1(\cdot)),\dots,(\alpha_m,T_m(\cdot))\}$
\end{algorithmic}
\end{algorithm}

With the return of Alg.~\ref{alg:samme}, given a testing data $\boldsymbol{x}_t$, a class label $\hat{y_t} \in \{1,2,\dots,K\}$ is assigned by $\argmax_{k}\sum^{M}_{m=1}\alpha_m\cdot\mathbb{I}(T_m(\boldsymbol{x}_t)=k)$.
\section{Problem Definition and Proposed Solution}
\subsection{Problem Definition}
\begin{figure}[htbp]
  \centering
    \includegraphics[width=0.5\textwidth]{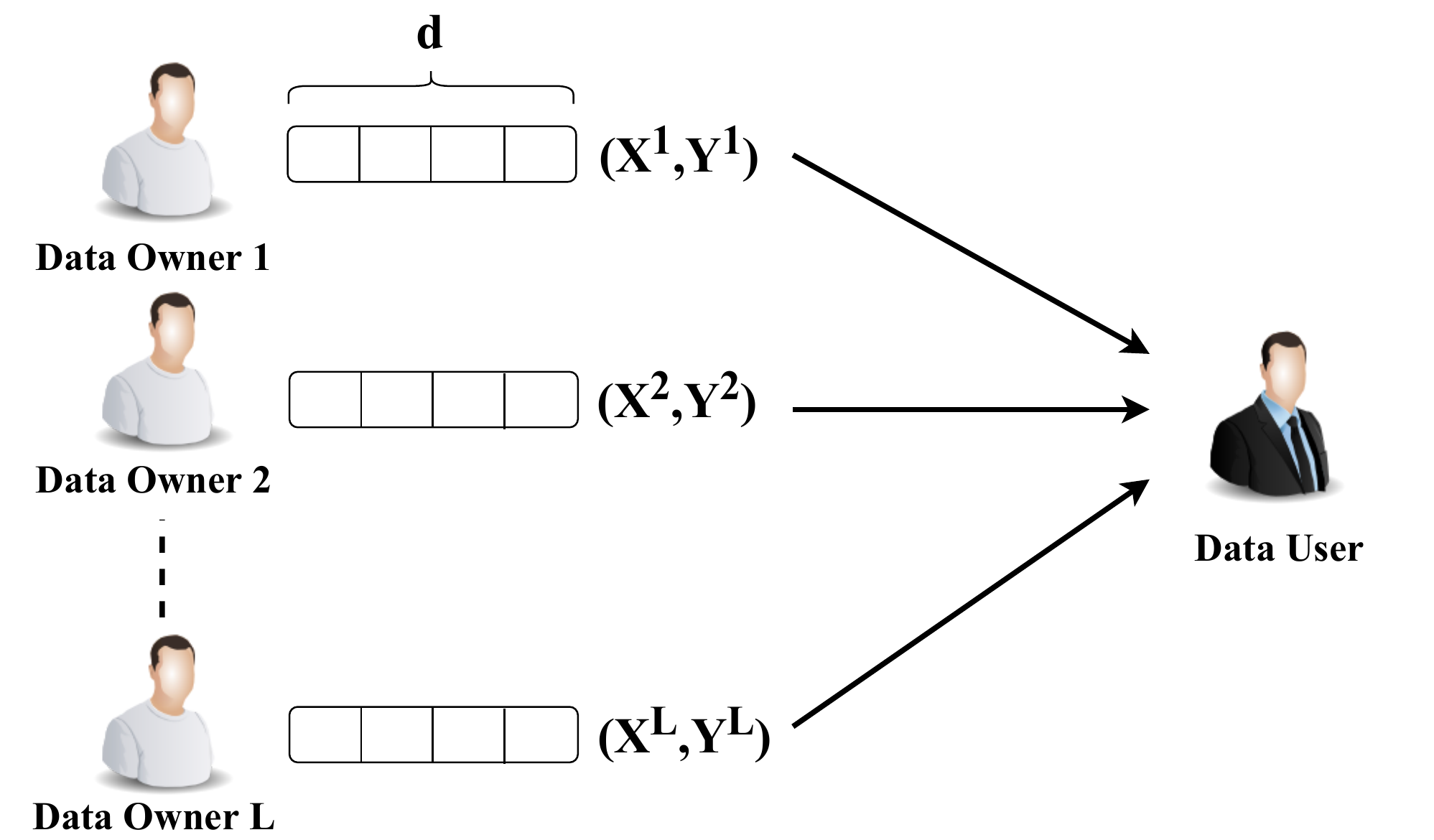}
  \caption{Problem Overview. Data owner $l$ holds a set of samples $(X^l,Y^l)$, $x^l_i \in \mathbb{R}^{d}$, and  $y^l_i \in \{1,2,\dots,K\}$ is the label associates with $x^l_i, i=1,2,\dots,N^l$ ; the untrustworthy data user would like to fit a boosted classifier with training samples from all $L$ data owners. The privacy of $x_i^l$ should be protected against the untrustworthy data user.}
  \label{fig:problem}
\end{figure}

In this paper, we are interested in developing a privacy-preserving boosting algorithm, As Fig.\ref{fig:problem} shows, we consider the following problem: \textit{Given $L$ data owners, each data owner $l$ holds a set of samples $(X^l,Y^l), x^l_i \in \mathbb{R}^{d}$, and $y^l_i \in \{1,2,\dots,K\}$ is the label associates with $x^l_i, i=1,2,\dots,N^l$; the untrustworthy data user would like to fit a boosted classifier with $(X^l,Y^l), l\in \{1,2,\dots,L\}$. Note in the paper, we assume that the data possessed by the data owner is represented as numerical values; for the categorical attribute that with $k$ distinct values, it can be transformed to $k$ binary attributes using one-hot encoding and then be proceeded accordingly.} The privacy constraint comes in two-folds at the data owner side, firstly, the data owner is not willing to share the value of individual sample to the data user; secondly, any inference of the individual data sample should be prevented from the intermediate communication messages. The symbols and notations used in this paper are summarized in Table.~\ref{tab:notation}.

\subsection{Threat Model}
In the problem, we assume that the data user is untrustworthy and the data owners are honest-but-curious, in which every data owner is obliged to follow the protocol, but intentionally likes to extend their knowledge during the execution of the protocol. \textit{And the value of the individual sample is what each party intends to acquire beyond the final classifier.} Thus an adversary could be either the data user, a participating data owner or an outside attacker, who intends to learn the value of data instances possessed by any data owner. It assumes that adversaries might have arbitrary background knowledge and there might be a collusion among them. The goal is to enforce the privacy of individual data instance while maintain the utility of the learned classifier. Furthermore, the untrustworthy data user could behave dishonestly, which would not compromise data owner's privacy with our solution, but will hurt the utility of the learned classifier. Therefore, it is of the data user's interest to correctly execute the algorithm. As such, our solution protects the privacy of the data instance from each data owner. Since the data owner is assumed to be honest-but-curious, data pollution attacks (e.g., data owners maliciously modify their inputs to bias the classifier learned by the data user) are beyond the scope of this paper.

\subsection{Privacy-Preserving Boosting}
We first provide an overview of our privacy-preserving boosting algorithm, which is run by the data owner and data user cooperatively. Recall that boosting algorithm is design to combine multiple base classifiers into a strong one, and each base classifier is trained by the samples with readjusted weights. Our proposed algorithm follows the same idea, while intends to prevent the leakage of the true value of individual training sample at the data owner. In general, the pseudo-code of the algorithm is given in Alg.~\ref{alg:pp_samme}:

\begin{algorithm}
\caption{Privacy-Preserving AdaBoosting}
\label{alg:pp_samme}
\begin{algorithmic}[1]
\STATE \textbf{Input}: $\{(X^1,Y^1),(X^2,Y^2),\dots,(X^L,Y^L)\}$.
\FOR{l=1 to L:}
\STATE Initialize the observation weights $w_i^l = 1/N^l, i=1,2,\dots,N^l$
\ENDFOR
\FOR{m=1 to M:}
\FOR{l=1 to L:}
\STATE Data owner $l$ prepares the local share $U^l$ and perturbs the local share to get $U^{l'}$.
\ENDFOR
\STATE Data user randomly selects a group of data owners, denoted as $H$ and collects the perturbed local share
\STATE $T_m \leftarrow Aggregate(U^{1'},\dots,U^{H'})$.

\STATE $WeightedError \leftarrow SecureSum(\sum^{N^1}_{i=1}w_i^1\mathbb{I}(y_i^1\neq T_m(\boldsymbol{x}_i^1)),\dots,\sum^{N^L}_{i=1}w_i^L\mathbb{I}(y_i^L\neq T_m(\boldsymbol{x}_i^L))$
\STATE $WeightsSum \leftarrow SecureSum(\sum^{N^1}_{i=1}w_i^1,\dots,\sum^{N^L}_{i=1}w_i^L)$
\STATE $err_m \leftarrow \frac{WeightedError}{WeightsSum}$
\STATE $\alpha_m \leftarrow log\frac{1-err_m}{err_m}+log(K-1)$
\FOR{l=1 to L:}
\STATE $w_i^l \leftarrow w_i^l\cdot exp(\alpha_m\cdot \mathbb{I}(y_i^l\neq T_m(\boldsymbol{x}_i^l))),i=1,2,\dots,N^l$
\ENDFOR
\STATE $WeightsSum \leftarrow SecureSum(\sum^{N^1}_{i=1}w_i^1,\dots,\sum^{N^L}_{i=1}w_i^L$
\FOR{l=1 to L:}
\STATE $w_i^l \leftarrow \frac{w_i^l}{WeightsSum}$
\ENDFOR
\ENDFOR

\RETURN $\{(\alpha_1,T_1(\cdot)),\dots,(\alpha_m,T_m(\cdot))\}$
\end{algorithmic}
\end{algorithm}

The whole algorithm consists of three stages: 1), line 6-8, each data owner prepares a local share and perturb it accordingly, the representation of the local share is different according to the type of the base classifier, and it will be further demonstrated in later section; 2), line 9-10, data owner sends the perturbed local share to the data user, and data user aggregates each to obtain the base classifier for the current round, then the base classifier is sent back to each data owner; 3), line 11-21, each data owner computes the weighted error rate and the sum of weights separately, then updates the weights of the local samples by collaborating with other data owners. Overall, line 6-21 completes a single round of the boosting algorithm, and line 23 outputs the final meta-classifier after such $m$ rounds. To compute the weighted error rates and sum of weights securely, data owners corroboratively run a secure sum computation protocol, like Shamir secret sharing\cite{kantarcioglu2004privacy,shamir1979share} in line 11,12 and 18. Similar MPC procedure has also been adopted in literature\cite{pathak2010multiparty}. In our paper, the reason to adopt MPC here is that the error rates and weights are aggregated statistics of the samples held by each data owner and they should be kept secret to minimize the potential information leakage, while such statistics are independent to the value of the individual sample and they are not necessarily be perturbed. Next, we give an example of computing the sum of weights, data owner 1 adds a random number $g^1$ to $\sum^{N^1}_{i=1}w_i^1$ and sends it to data owner 2. Data owner 2 adds a random number $g^2$ and $\sum^{N^2}_{i=1}w_i^2$ to the received value from data owner 1 and sends it to data owner 3. The process continues till data owner $L$ adds a random number $g^L$ and $\sum^{N^L}_{i=1}w_i^L$ to received value from data owner $L-1$ and sends it to data owner 1. Data owner 1 subtracts $g^1$ from the received value from data owner $L$ and sends back the result to data owner 2. Data owner 2 subtracts $g^2$ and sends back the result to data owner 3. Process continues till data owner $L$ subtracts $g^L$. Finally, data owner $L$ gets the total sum of weights and distributes to all other data owners.

\subsection{Local Share Perturbation}
In the proposed algorithm, each data owner perturbs the local share and then sends the perturbed share to the data user to fit the base classifier. In our definition, data user should not be able to infer any individual sample from the received share. As mentioned previously, according to the definition of LDP, by seeing a perturbed data instance, it can't distinguish it from the true data instance with a high confidence (controlled by the parameter $\epsilon$). Thus to protect the privacy of the individual data instance, the randomization method which satisfies $\epsilon$-LDP is adopted. In the meanwhile, one challenge is also raised as how the utility of the base classifier be maintained by building upon the perturbed data instances. To the best of our knowledge, the statistical information of the perturbed dataset could be maintained with an error bound\cite{wang2017locally,duchi2018minimax}, which gives us a way to build the base classifier. Be more specifically, the statistical information (e.g., the estimated mean) of the perturbed local shares are used to build the base classifier, thus the utility of the classifier could be guaranteed. To achieve the goal, we first introduce the method\cite{wang2019collect} that is used to perturb the numerical attributes and show the error bound for the mean estimation. For ease of explanation, the individual sample held by a data owner, $\boldsymbol{x}\in \mathbb{R}^{d}, \boldsymbol{x}=[x_1,x_2,\cdots ,x_d]$, is assumed to be perturbed here.

\begin{algorithm}
\caption{Piecewise Mechanism for One-Dimensional Numeric Data \cite{wang2019collect}}
\label{alg:pm_1d}
\begin{algorithmic}[1]
\STATE \textbf{Input}: tuple $x_i\in [-1,1]$ and privacy parameter $\epsilon$
\STATE \textbf{Output}: tuple $x_i' \in [-\Delta,\Delta]$
\STATE $\Delta\leftarrow \frac{e^{\epsilon/2}+1}{e^{\epsilon/2}-1}$
\STATE $\phi_{left}(x_i) \leftarrow \frac{\Delta+1}{2}\cdot x_i - \frac{\Delta-1}{2}$
\STATE $\phi_{right}(x_i) \leftarrow \phi_{left}(x_i)+\Delta-1$
\STATE Sample $v$ uniformly at random from $[0,1]$;
\IF{$v< \frac{e^{\epsilon/2}}{e^{\epsilon/2}+1}$}
\STATE Sample $x_i'$ uniformly at random from $[\phi_{left}(x_i),\phi_{right}(x_i)]$
\ELSE
\STATE Sample $x_i'$ uniformly at random from $[-\Delta,\phi_{left}(x_i)] \cup [\phi_{right}(x_i),\Delta]$
\ENDIF
\RETURN $x_i'$
\end{algorithmic}
\end{algorithm}

Alg.~\ref{alg:pm_1d} takes a single numerical value $x_i\in [-1,1]$ as input and returns a perturbed value $x_i' \in [-\Delta,\Delta]$, where it confines $x_i'$ to a relatively small domain (i.e., $x_i' \in [-\Delta,\Delta]$) while it also allows $x_i'$ to be close to $x_i$ (i.e., $x_i' \in [\phi_{left}(x_i),\phi_{right}(x_i)]$) with reasonably large probability (i.e., $\frac{e^{\epsilon/2}}{e^{\epsilon/2}+1}$). For the mean estimation, with at least $1-\beta$ probability, it can be shown that $|\frac{1}{n}\sum_{i=1}^nx_i'-\frac{1}{n}\sum_{i=1}^nx_i| = O(\frac{\sqrt{log(1/\beta)}}{\epsilon\sqrt{n}})$, which is an error bound with the incurred noise and is asymptotically optimal\cite{duchi2018minimax}. In the case that the input domain is not within the range $[-1,1]$, namely, $x_i\in [-t,t], t>0$, the data owner computes $x_i^\ast = \frac{x_i}{t}$ and perturbs $x_i^\ast$ with Alg.~\ref{alg:pm_1d}, then the data owner releases $x_i^{\ast\prime} \cdot t$, where $x_i^{\ast\prime}$ is the perturbed output of $x_i^\ast$. The assumption that $t$ is public information is commonly used in the literature\cite{duchi2018minimax}.
\begin{algorithm}
\caption{Piecewise Mechanism for Multi-Dimensional Numeric Data\cite{wang2019collect}}
\label{alg:pm_md}
\begin{algorithmic}[1]
\STATE \textbf{Input}: tuple $\boldsymbol{x}\in [-1,1]^d$ and privacy parameter $\epsilon$
\STATE \textbf{Output}: tuple $\boldsymbol{x}' \in [-d\cdot \Delta,d\cdot \Delta]^d$
\STATE $\boldsymbol{x}'\leftarrow <0,0,\dots,0>$
\STATE $k \leftarrow max\{1,min\{d,\lfloor \frac{\epsilon}{2.5} \rfloor \}\}$
\STATE Sample $k$ values uniformly without replacement from $\{1,2,\dots,d\}$
\FOR{each sampled attribute $j$}
\STATE $x_j' = \frac{d}{k}Alg.3(x_j,\frac{\epsilon}{k})$
\ENDFOR
\RETURN $\boldsymbol{x}'$
\end{algorithmic}
\end{algorithm}

when $\boldsymbol{x}$ is a multi-dimensional vector, namely, $d>1$, Alg.~\ref{alg:pm_md} is adopted, which selects $k$ attributes, $k<d$ and perturbs each with Alg.~\ref{alg:pm_1d}. The intuition behind Alg.~\ref{alg:pm_md} is to reduce the amount of the noise in the estimated mean of $\boldsymbol{x}$, where $k$ is the value that chosen to minimize the worst-case noise variance. Comparing to the solution that perturbing $d$ attributes with Alg.~\ref{alg:pm_1d}, where each attribute is evenly given a privacy budget of $\frac{\epsilon}{d}$ and the total amount of noise in the mean estimation is in $O(\frac{d\sqrt{logd}}{\epsilon\sqrt{n}})$, which is super-linear to $d$; it can be proven that, with Alg.~\ref{alg:pm_md}, $E[max_{j\in[1,d]}|\frac{1}{n}\sum_{i=1}^nx_{i,j}'-\frac{1}{n}\sum_{i=1}^nx_{i,j}|] = O(\frac{\sqrt{dlog(d/\beta)}}{\epsilon\sqrt{n}})$ with at least $1-\beta$ probability, and it is still asymptotically optimal, where the full proof is referred to the original work\cite{wang2019collect}.

\subsubsection{Privacy Budget Consumption}
With Alg.\ref{alg:pm_md}, we now take a look back of our proposed privacy-preserving boosting algorithm and analyze the privacy and utility trade-off. In Alg.~\ref{alg:pp_samme}, the final classifier is a combination of $m$ base classifiers with different voting weights, which are built after $m$ rounds of computation. In the non-private scenario, the data owner is able to participate in all $m$ rounds computation. However, in the LDP setting, the privacy budget is affected by the repeating involvements of a data owner. More specifically, suppose the local share released by the data owner in the $i$-th round satisfies $\epsilon_i$-differential privacy. By the composition property of Differential Privacy\cite{mcsherry2007mechanism}, if $\epsilon$-DP is required to be enforced for the data owner's data, then it needs that $\sum_{i}^m\epsilon_i \leq \epsilon$. Considering that the data owner participates all $m$ rounds, it becomes $\epsilon_i=\epsilon/m$. Then the amount of the noise contributed by each data owner becomes $O(\frac{m\sqrt{dlogd}}{\epsilon})$, which linearly depends on $m$. Thus to reduce the injected noise, following the suggestion in the existing studies \cite{hamm2015crowd,duchi2018minimax}, each data owner only participates in at most one round. In particular, as line 9-10 of Alg.~\ref{alg:pp_samme} shows, in each round, the data user uniformly at random selects a group $H$ of data owners, and asks each of them to submit the perturbed local share. Then the data user aggregates $|H|$ perturbed local shares in each round, which results the amount of the noise in $O(\frac{\sqrt{dlogd}}{\epsilon\sqrt{|H|}})$.

\subsection{Building Base Classifier}
After giving an overview of the privacy-preserving boosting algorithm and the perturbation methods, we now turn to the exact classifier that to be fit. For a long time, Boosting Decision Tree (BDT) has been used to demonstrate the effectiveness of boosting, while the idea is actually applied to other types of classifiers. In the subsection, we give three examples of classifiers that could be supported and discuss how to build such classifiers with different types of local shares. Each final classifier  consists of $m$ base classifiers with different weights, and the procedure to build each base classifier keeps the same but with adjusted sample weights.

\subsubsection{Local Samples}
A straightforward approach to build the base classifier is to contribute the data samples directly by each data owner, then the data user collects such input and build the classifier at a centralized site. However, to protect the privacy, the data samples should be perturbed before releasing. Due to the noise injected to each data instance, only the statistical information could be well estimated, such as frequency and mean estimation, which makes only certain types of classifiers could be supported. Nearest Centroid Classifier (NCC) is one such classifier could be built upon the perturbed data samples, where the observation is assigned with the label of the class of training samples whose centroid is nearest to that observation. For data owner $l$, $\boldsymbol{x}_i^\ast = w_i\cdot \boldsymbol{x}_i$ is the weighted sample, $\boldsymbol{x}_i^{\ast\prime}$ is the perturbed output of $\boldsymbol{x}_i^\ast$ by Alg.\ref{alg:pm_md} and $(\boldsymbol{x}_i^{\ast\prime},y_i)$ is submitted to the data user. For the data user, the centroid of class $C_k$ is computed as $\boldsymbol{\mu}_k = \frac{1}{|C_k|}\sum_{i\in C_k}\boldsymbol{x}_i^{\ast\prime}$. Given an observation $\boldsymbol{x}_t$, the distance to each class centroid is computed and then the class label of closet centroid, $\hat{y}_t=\argmin_{k}||\boldsymbol{\mu}_k-\boldsymbol{x}_t||_\ell$, is assigned to $\boldsymbol{x}_t$.

\subsubsection{Local Classifier}
Contributing the perturbed data samples is a straightforward way, but the supported classifiers are limited due to that only the mean of the data samples could be well estimated. As an alternative, each data owner could locally compute an optimal classifier using their own samples. Instead of submitting the data samples, the individual classifier is then perturbed and contributed to the data user. More specifically, assuming the classifier is trained by optimizing the loss function $\mathcal{L}(\cdot)$ that maps a $d$-dimensional parameter vector $\boldsymbol{\theta}$ into a real number, where the parameter vector $\hat{\boldsymbol{\theta}^l}$ computed by data owner $l$ is identified as:
\begin{equation}
\hat{\boldsymbol{\theta}^l}  = \argmin_{\boldsymbol{\theta}^l}[\frac{1}{N^l}(\sum_{i=1}^{N^l}\mathcal{L}(\boldsymbol{\theta}^l;w_i^l\cdot\boldsymbol{x_i}^l;y_i^l)]
\end{equation}
Similarly, $\hat{\boldsymbol{\theta}^l}'$ is the perturbed output of $\hat{\boldsymbol{\theta}^l}$ by Alg.\ref{alg:pm_md}. Then the data user collects $L$ perturbed parameter vectors and obtains the final parameter vector:
\begin{equation}
\hat{\boldsymbol{\theta}} = \frac{1}{L}\sum_{l}\hat{\boldsymbol{\theta}^l}' 
\end{equation}

It is not hard to see that submitting the local classifiers is a superior solution comparing to submitting the training samples, while the accuracy of the estimation is affected by the number of collected samples. Given $L$ data owners, there are $L$ local classifiers and the estimated error of the aggregation is bounded by $O(\frac{\sqrt{dlogd}}{\epsilon\sqrt{L}})$. However, each data owner may hold more than one training sample, assuming the total amount of the samples is $n$, when $n\gg L$, the estimated error will be far less than aggregation of $L$ parameters.

\subsubsection{Local Statistic}
Another type of base classifiers is built by the statistic information of the data samples. In this case, rather than individual samples or the local classifiers, the data owner computes the statistic of the data samples and submits to the data user. BDT is one such classifier could be built in this way, and we explain the fitting procedure here. BDT is one type of Decision Tree (DT) classifier, which makes the decision by examining a sequence of conditions and it is represented as a tree from the root node to the leaf node. The internal node of the tree represents a condition, and the corresponding value in the observation decides which branch to proceed, till the leaf nodes provides the final decision. Thus building a DT from the root to the leaves needs to pick the proper attribute for each internal node and the process repeats till the tree stop growing. The best attribute to split is determined by the impurity measurement before and after split over that attribute. The impurity measurements usually include information gain\cite{quinlan1986induction}, Gini index\cite{breiman2017classification} and misclassification error\cite{tan2018introduction}, and the former two are commonly used in the DT. Comparing to DT that usually resulted in multiple levels, BDT takes advantage of the decision stumps of one depth and the final classifier makes the final decision by considering the weighted votes of all stumps. And we choose the measurement of misclassification error in this paper, as the previous study\cite{bhaduri2008distributed} shows that the misclassification error fits well in the distributed environment. To summarize, the base classifier in BDT is the decision stump which has a binary branch and to find the best attribute to be split on this stump, we use the misclassification error as the criteria. Next, we describe how to privately find the best attribute based on the misclassification error and to build the decision stump.

For ease of illustration, assuming $A^j$ is a binary attribute, and there are two classes in the learning examples, the cross table of $A^j$ is a statistical summary and is represented as $S^j = \Big(\begin{matrix}
    s_{0,0}^j &  s_{0,1}^j \\
    s_{1,0}^j &  s_{1,1}^j
  \end{matrix}\Big)$, where $s_{0,1}^j$ denotes the weighted number of examples with $A^j=0$ and $Class = 1$. In short, for all samples with $A^j=0$ and belong to Class 1, instead of adding the counts, $s_{0,1}^j$ is the added weights of the corresponding samples. The misclassification error impurity of the $A^j=0$ is defined as $1-max\{s_{0,0}^j,s_{0,1}^j\}/s_0^j$, where $s_0^j$ is the weighted number of samples that $A^j=0$. Then it can be shown that for each attribute, maximizing its gain is equivalent to minimizing the weighted sum of impurities\cite{bhaduri2008distributed}:
  \begin{multline}
  \argmin_j\Big\{ \frac{s_0^j}{|S^j|}\Big[ 1-\frac{max\{s_{0,0}^j,s_{0,1}^j\}}{s_0^j} \Big] +\\ \frac{s_1^j}{|S^j|}\Big[ 1-\frac{max\{s_{1,0}^j,s_{1,1}^j\}}{s_1^j} \Big]\Big\} \\ 
  \iff \argmax_j\{ |s_{0,0}^j-s_{0,1}^j|+|s_{1,0}^j-s_{1,1}^j| \}
  \end{multline}
Thus the best attribute $A^{best}$ is the one that has the minimum classification error, namely, $|s_{0,0}^{best}-s_{0,1}^{best}|+|s_{1,0}^{best}-s_{1,1}^{best}| \geq |s_{0,0}^j-s_{0,1}^j|+|s_{1,0}^j-s_{1,1}^j|, \forall A^j\neq A^{best}$. In our problem, the data user aggregates the cross tables of all attributes from the data owners and to determine the best attribute. More specifically, each data owner prepares two values for each attribute, $\{s_{0,0}^j-s_{0,1}^j,s_{1,0}^j-s_{1,1}^j\}, j=1,2,\cdots,d$ and submits the perturbed values to the data user, e.g., $\{(s_{0,0}^j-s_{0,1}^j)',(s_{1,0}^j-s_{1,1}^j)'\}$. Upon receiving the statistics from $L$ data owners, the data user sums up the values and computes $\{|\sum_l[(s_{0,0}^j-s_{0,1}^j)']|,|\sum_l[(s_{1,0}^j-s_{1,1}^j)']|\}$. Furthermore, since it is only required to decide the attribute that has minimum classification error, and the actual error of each attribute is not necessarily needed. Thus the data user could use the mean estimation to determine the best attributes, namely, $\{|\frac{\sum_l[(s_{0,0}^j-s_{0,1}^j)']}{l}|,|\frac{\sum_l[(s_{1,0}^j-s_{1,1}^j)']}{l}|\}$. Thus the data user determines the best attributes which has the minimum classification error and broadcasts such attribute to all data owners. Knowing the best attribute, each data owner reports the majority of class labels for each branch to the data user (recall it is a binary attribute), thus the base classifier is determined by the best attribute and the majorities class labels for each branch.

\section{Experiment}
In this section, we evaluate the proposed privacy-preserving boosting algorithm through three types of classifiers, Logistic Regression (LR), Nearest Centroid Classifier (NCC), Boosted Decision Tree (BDT); each classifier corresponds to one subsection of Section 3.5. The utility of the learned classifiers are assessed in terms of the prediction capacity over three public datasets, the MNIST, Fashion-MNIST\footnote{https://github.com/zalandoresearch/fashion-mnist}, and IPUMS dataset. The description of the datasets are given below, 
\begin{figure*}[t]
\centering
\begin{subfigure}[t]{1\textwidth}
\begin{subfigure}[t]{.32\textwidth}
\centering
\includegraphics[width=1\textwidth]{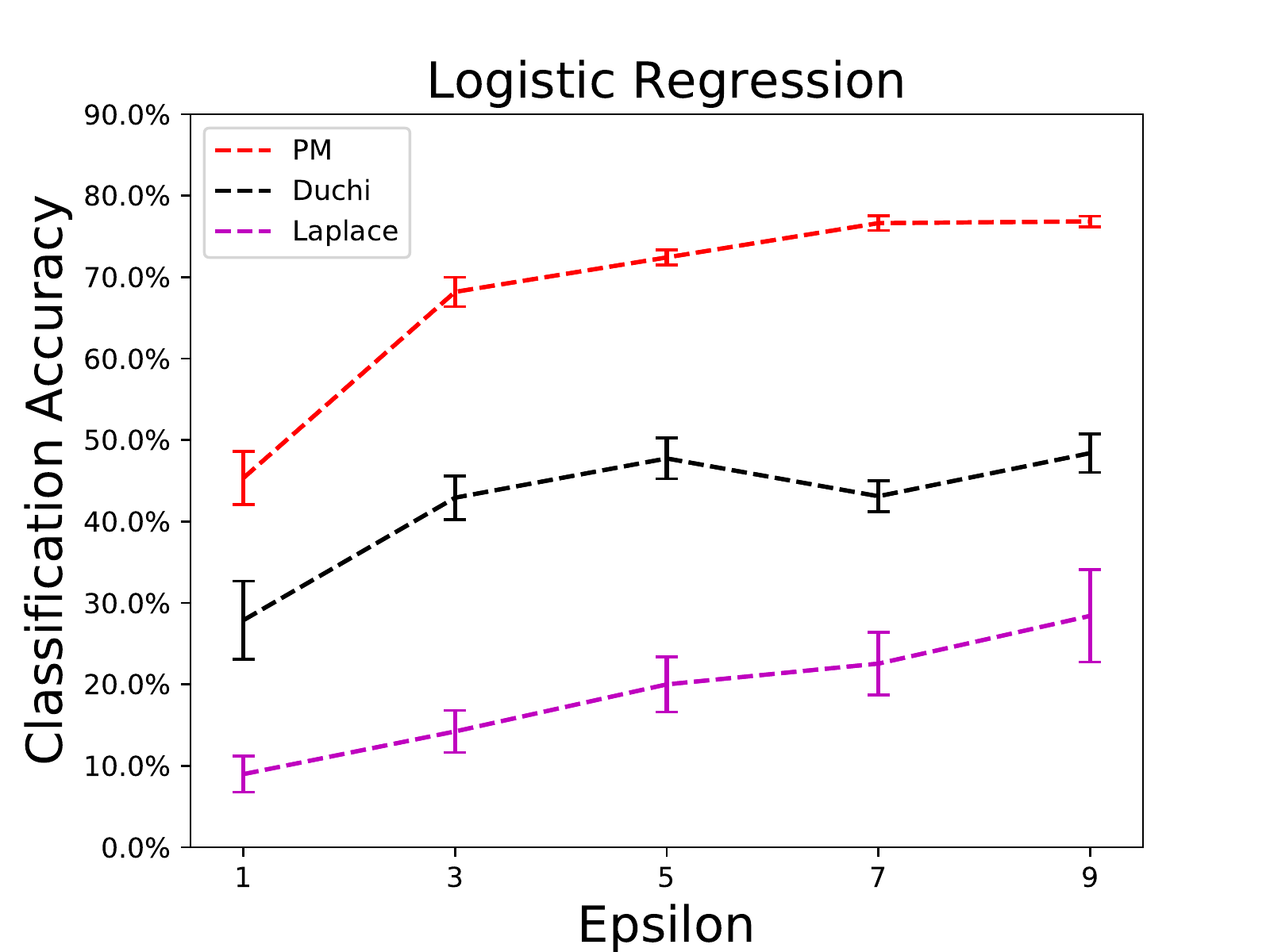}
\caption{MNIST}
\end{subfigure}
\begin{subfigure}[t]{.32\textwidth}
\centering
\includegraphics[width=1\textwidth]{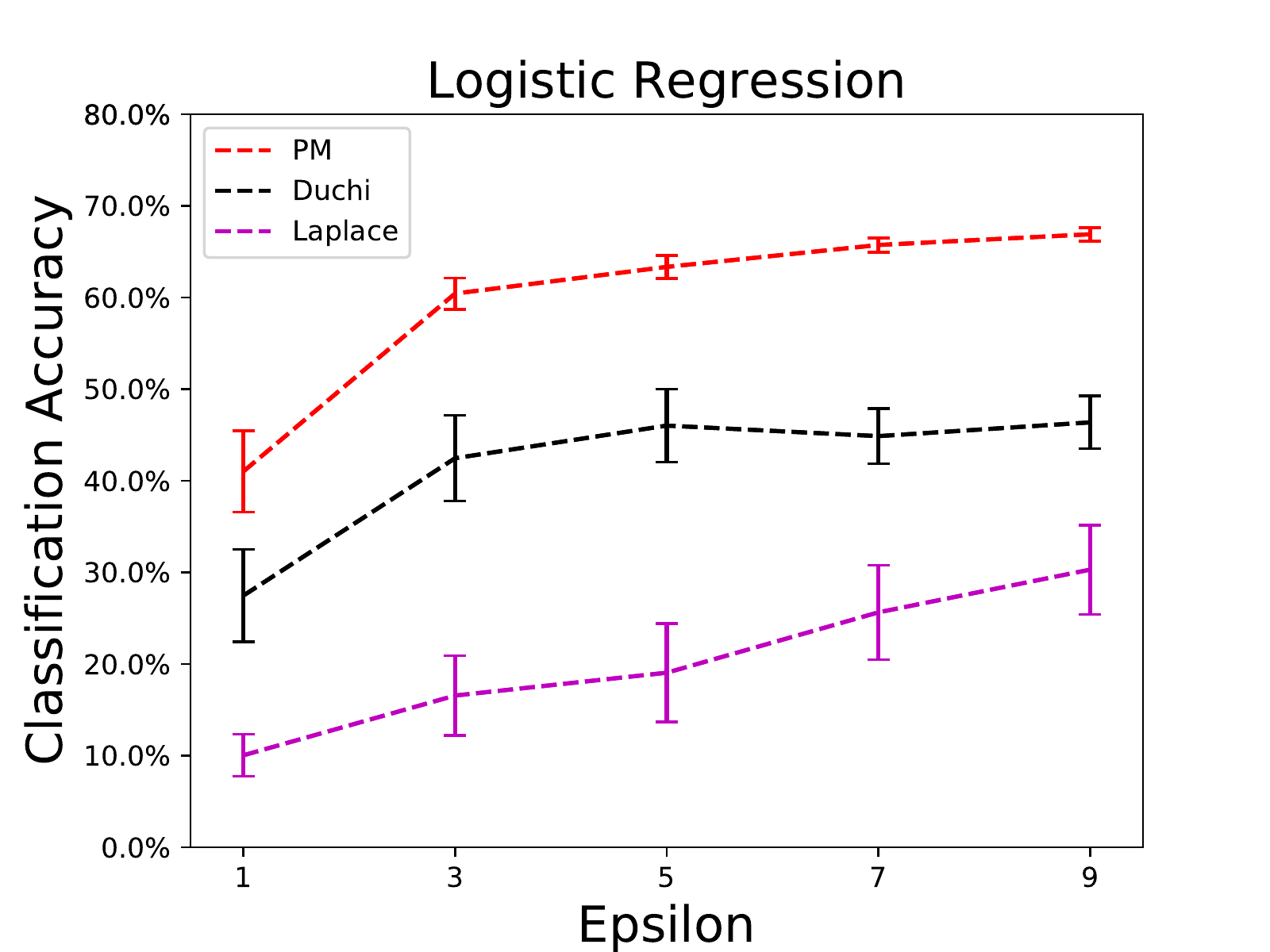}
\caption{Fashion-MNIST}
\end{subfigure}
\begin{subfigure}[t]{.32\textwidth}
\centering
\includegraphics[width=1\textwidth]{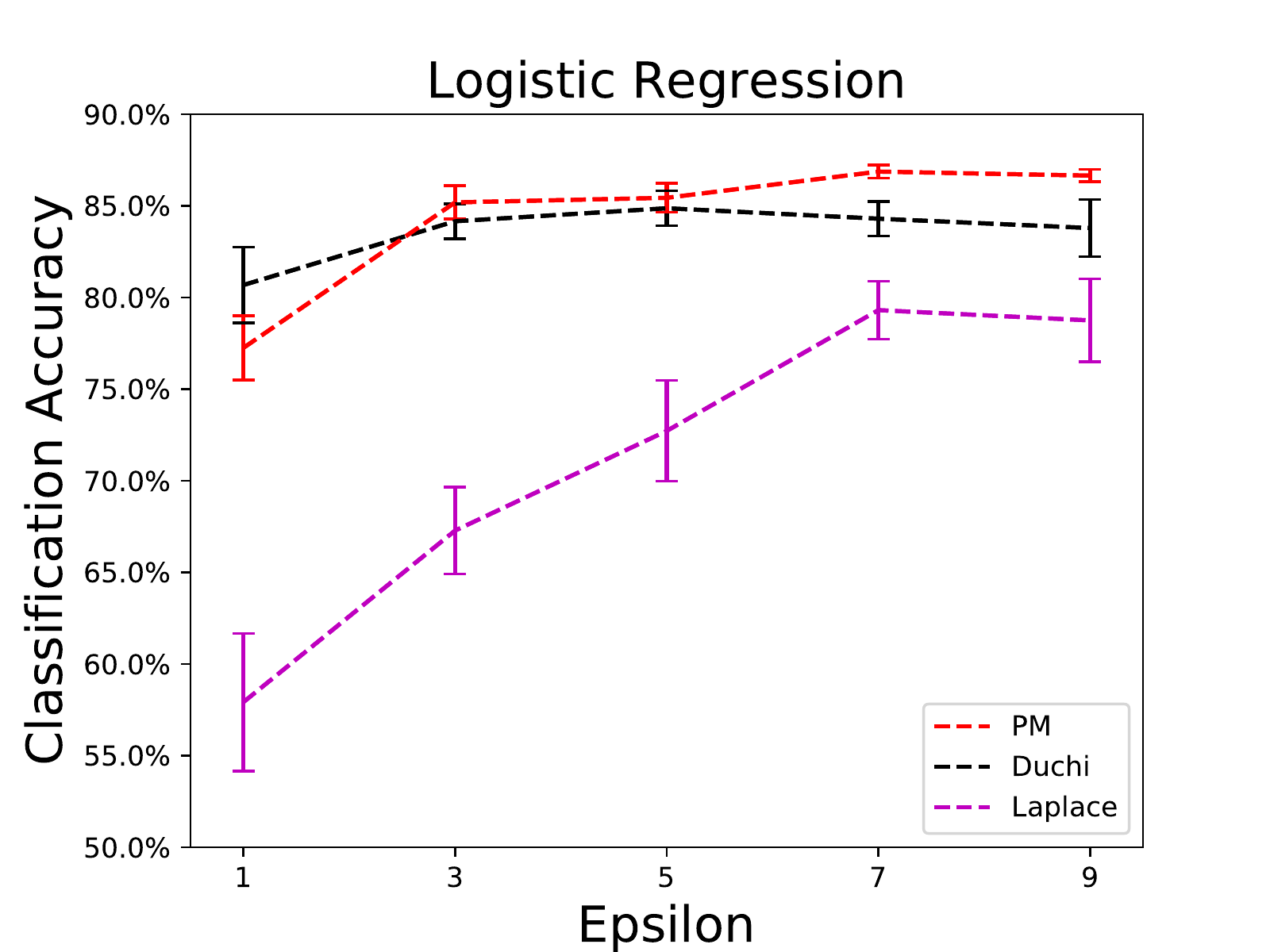}
\caption{Brazil}
\end{subfigure}
\end{subfigure}
\medskip
\caption{LR classification accuracy v.s. privacy budgets. Each subfigure shows the result for one dataset in terms of Laplace, Duchi and PM perturbation methods respectively. For MNIST and Fashion-MNIST, the dimensionality of the datasets is 50, and there are 500 data owners. For Brazil, the dimensionality of the dataset is 67, and there are $10^5$ data owners.}
\label{fig:lr}
\end{figure*}

\begin{table*}[t]
\centering
\caption{MSE of the mean estimation of the collected parameters of LR classifier in terms of Laplace, Duchi and PM respectively.}
\begin{tabular}{|c|c|c|c|c|c|c|}
\hline
\multicolumn{1}{|l|}{} & \diagbox{Perturbation}{$\epsilon$}      & 1.0            & 3.0            & 5.0            & 7.0           & 9.0             \\ \hline
\multirow{3}{*}{MNIST} & Laplace         & 27.4852 + 0.7686 & 2.9896 + 0.1591 & 1.0848 + 0.0801 & 0.5387 + 0.0302 & 0.3333 + 0.0148 \\ \cline{2-7} 
                             & Duchi & 0.6149 + 0.0367 & 0.1566 + 0.0100 & 0.1300 + 0.0073 & 0.1347 + 0.0086 & 0.1350 + 0.0075 \\ \cline{2-7} 
                             & PM &  \textbf{0.5104 + 0.0417} &   \textbf{0.0381 + 0.0032}    &  \textbf{0.0299 + 0.0021}    &  \textbf{0.0139 + 0.0007}    & \textbf{0.0129 + 0.0013}  \\ \hline

\multirow{3}{*}{Fashion-MNIST} & Laplace  & 26.6123 + 0.8602 & 3.1190 + 0.1216 & 1.0818 + 0.0787 & 0.5538 + 0.0332 & 0.3353 + 0.0130  \\ \cline{2-7} 
                             & Duchi & 0.6189 + 0.0203 & 0.1614 + 0.0077 & 0.1334 + 0.0126 & 0.1389 + 0.0066 & 0.1344 + 0.0099  \\ \cline{2-7} 
                             & PM        &  \textbf{0.4861 + 0.0235}     &   \textbf{0.0361 + 0.0023}   &  \textbf{0.0280 + 0.0014}     &  \textbf{0.0134 + 0.0015}     & \textbf{0.0120 + 0.0006}  \\ \hline
\multirow{3}{*}{Brazil} & Laplace & 11.9691 + 1.3442 & 1.2872 + 0.1661 & 0.5361 + 0.0904 & 0.2347 + 0.0295 & 0.1658 + 0.0286 \\ \cline{2-7} 
                             & Duchi & \textbf{0.1616 + 0.0227} & 0.0436 + 0.0070 & 0.0377 + 0.0075 & 0.0400 + 0.0059 & 0.0372 + 0.0027 \\ \cline{2-7} 
                             & PM &  0.2050 + 0.0200 &   \textbf{0.0100 + 0.0021}    & \textbf{0.0105 + 0.0024}    &  \textbf{0.0034 + 0.0012}    & \textbf{0.0043 + 0.0006} \\ \hline

\end{tabular}
\label{tab:mse_lr}
\end{table*}

\textbf{The MNIST dataset} contains $28 \times 28$ grayscale images of handwritten digits from 0 to 9, which has 60,000 samples for training and 10,000 samples for testing.

\textbf{The Fashion-MNIST dataset} contains the grayscale article images that each sample associates with a label from 10 classes. The dataset is intended to replace the overused MNIST dataset, and shares the same image size and structure of training and testing splits as the MNIST dataset.

\textbf{The Brazil dataset} contains census records of Brazil between 1970 and 2010 that from the Integrated Public Use Microdata Series\cite{ipums} (IPUMS). The dataset has 12M records and 10 attributes, 2 of them are numerical (e.g., age and totalIncome) and the rest of them are categorical. In the experiment, the totalIncome is used as the dependent variable and is converted to binary attribute by mapping the value larger than mean value to 1, and 0 otherwise\cite{wang2019collect}. The categorical attributes with k distinct values are transformed to k binary attributes as well. After the transformation, the dimensionality of the dataset becomes 67.

For better representation, the MNIST and Fashion-MNIST dataset are preprocessed with the Histogram of Oriented Gradients (HOG) descriptors, where the descriptor provides the histograms of directions of gradients that are used to capture the edges and corners of an image, which results in a fixed length of floating vector. Then a Bag of Visual Words (BOVW) model is further associated with the HOG descriptors, where the K-Means clustering is performed over the generated descriptors and the center of each cluster is used as the visual dictionary’s vocabularies. Finally, the frequency histogram is made from the visual vocabularies and results in a fixed length of vector. By this way, it is flexible to tune the number of visual vocabularies of both datasets and provides the representation with various dimensionalities.

For a comprehensive study, we further compare to two other existing numerical attributes perturbation method, Laplace mechanism\cite{dwork2006calibrating} and Duchi et al.'s mechanism. The Laplace mechanism perturbs the single numerical value by adding a random noise $Lap(0,2/\epsilon)$, in which $Lap(\mu,\lambda)$ follows a Laplace distribution that has 0 mean and scale $\lambda$. And it can be seen that the estimated mean of $n$ perturbed values is unbiased and the error is in $\frac{1}{\epsilon\sqrt{n}}$. However, to perturb multiple attributes, each attribute evenly split the total budget $\epsilon$, which is assigned $\epsilon/d$ individually and the error of the estimated mean of the multi-dimensional vector is super-linear to $d$, which will be excessive given a large $d$. Duchi et al.'s solution to perturb multiple numerical attributes\cite{duchi2018minimax} is also asymptotically optimal, but has a larger constant than Alg.~\ref{alg:pm_md}, and the effect in terms of the prediction capacity is demonstrated in the experiments. For consistency, in the figures of this section, \textit{Laplace} is referred to the Laplace mechanism, \textit{Duchi} is referred to the Duchi et al.'s mechanism and \textit{PM} is referred to Alg.~\ref{alg:pm_md}.

\begin{figure*}[ht!]
\centering
\begin{subfigure}[t]{1\textwidth}
\begin{subfigure}[t]{.32\textwidth}
\centering
\includegraphics[width=1\textwidth]{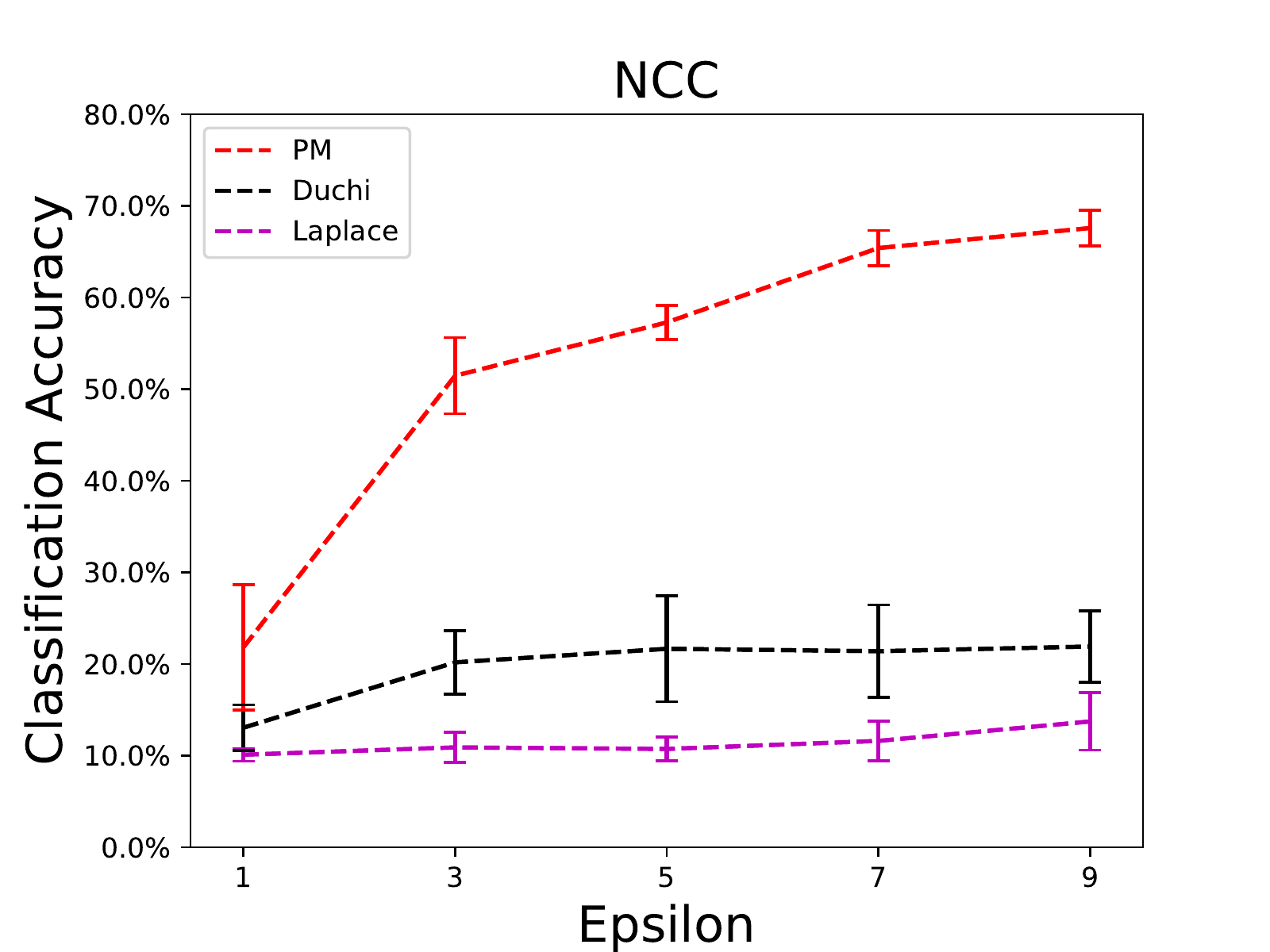}
\caption{MNIST}
\end{subfigure}
\begin{subfigure}[t]{.32\textwidth}
\centering
\includegraphics[width=1\textwidth]{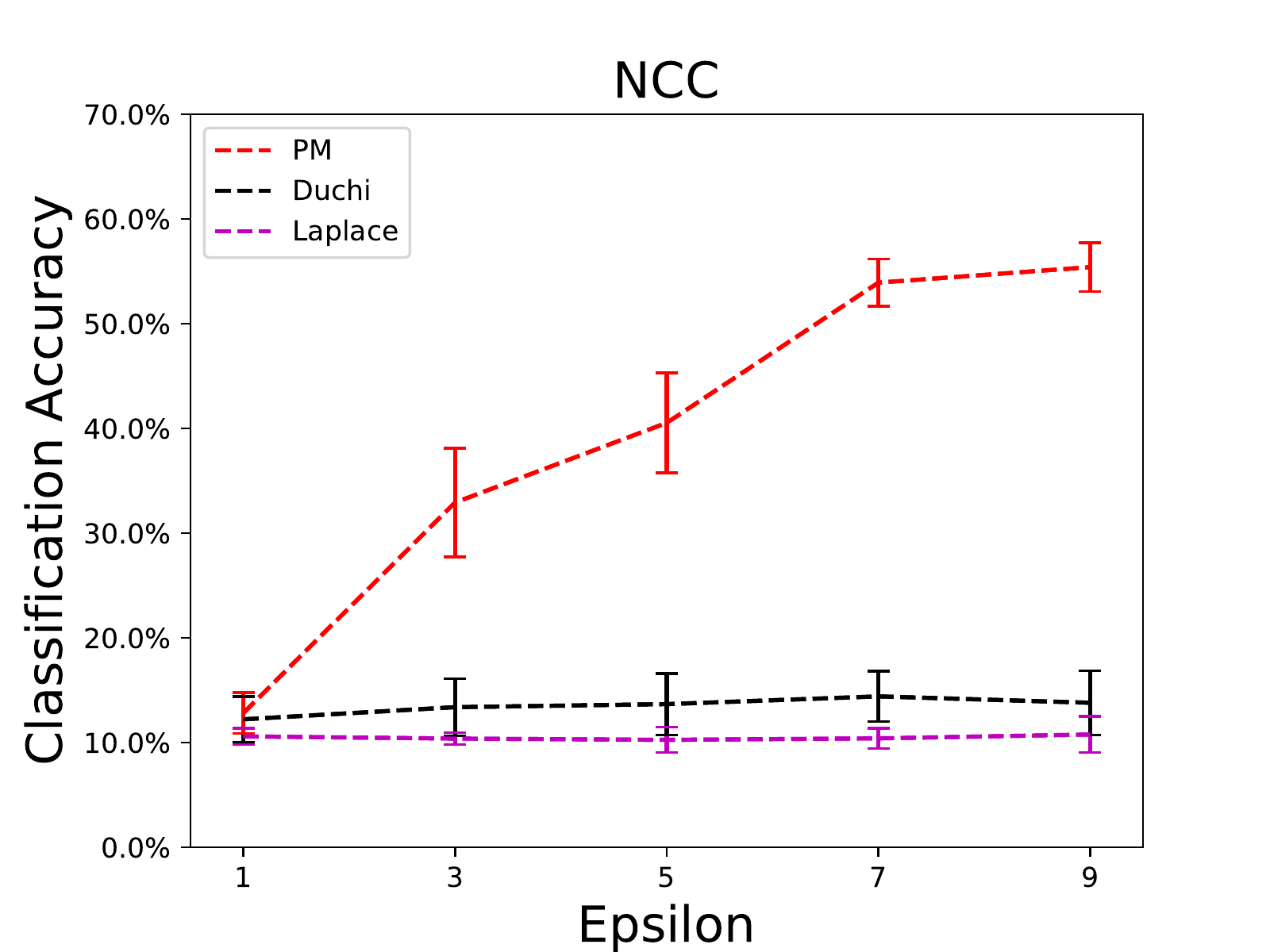}
\caption{Fashion-MNIST}
\end{subfigure}
\begin{subfigure}[t]{.32\textwidth}
\centering
\includegraphics[width=1\textwidth]{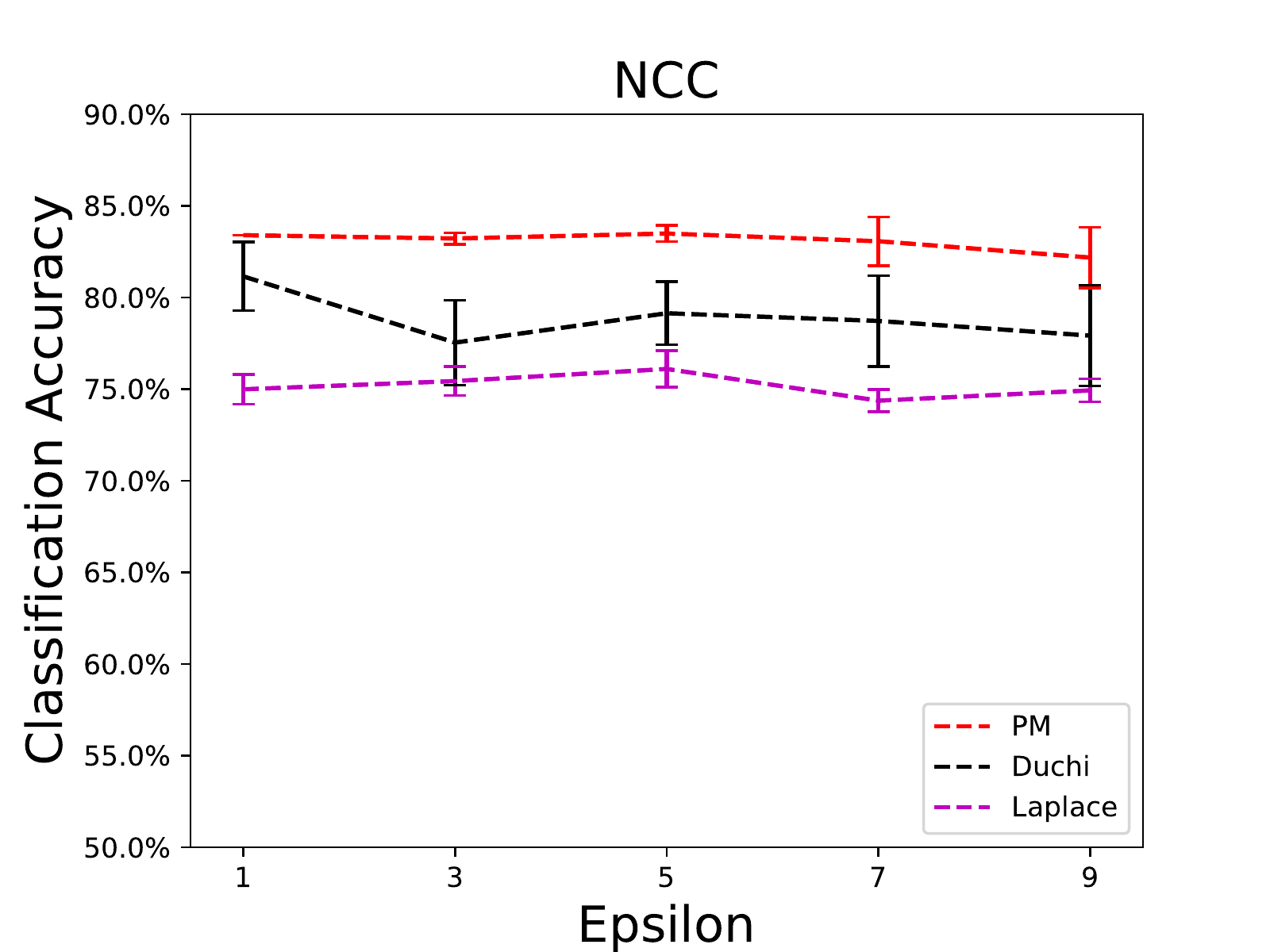}
\caption{Brazil}
\end{subfigure}
\end{subfigure}
\medskip
\caption{NCC classification accuracy v.s. privacy budgets. For MNIST and Fashion-MNIST, the dimensionality of the datasets is 50, and there are 10,000 samples collected in total. For Brazil, the dimensionality of the dataset is 67, and there are $10^6$ samples collected in total.}
\label{fig:ncc}
\end{figure*}

\subsection{Logistic Regression}

We first present the evaluation result regarding LR implementation. In the experiment, the dataset is randomly shuffled and evenly split into $L$ trunks to simulate $L$ data owners. Then a local LR classifier is built with each trunk of data and the parameter of the local classifier is perturbed by Alg.~\ref{alg:pm_md} to satisfy $\epsilon$-LDP. The base classifier is attained by taking the mean value of a group of the perturbed classifiers and the prediction capacity is evaluated. Fig.~\ref{fig:lr} displays the LR accuracy for all three datasets. Each subfigure compares three perturbation methods, the horizontal axis specifies privacy budget $\epsilon$, and the vertical axis shows the classification accuracy. From the figure, for all three perturbation methods, it can be seen that the accuracy improves as $\epsilon$ increases in general, as the incurred noise decreases, and PM among all three perturbation methods achieves the highest accuracy. The reason is that PM has the smallest error in the aggregation of the classifier parameters, and Table.~\ref{tab:mse_lr} presents the corresponding MSE of the estimated mean of the parameters. Comparing to Laplace and Duchi, PM has the smallest MSE across all privacy budgets except one setting, and the magnitude of the MSE difference reflects the pattern in Fig.~\ref{fig:lr}. For instance, for the MNIST and Fashion-MNIST datasets, the MSE of PM is one order of magnitude smaller than Duchi, and two order of magnitude smaller than Laplace, which shows the good utility of the learned base classifier.

\begin{figure}[h]
\begin{minipage}[t]{0.45\linewidth}
    \includegraphics[width=\linewidth]{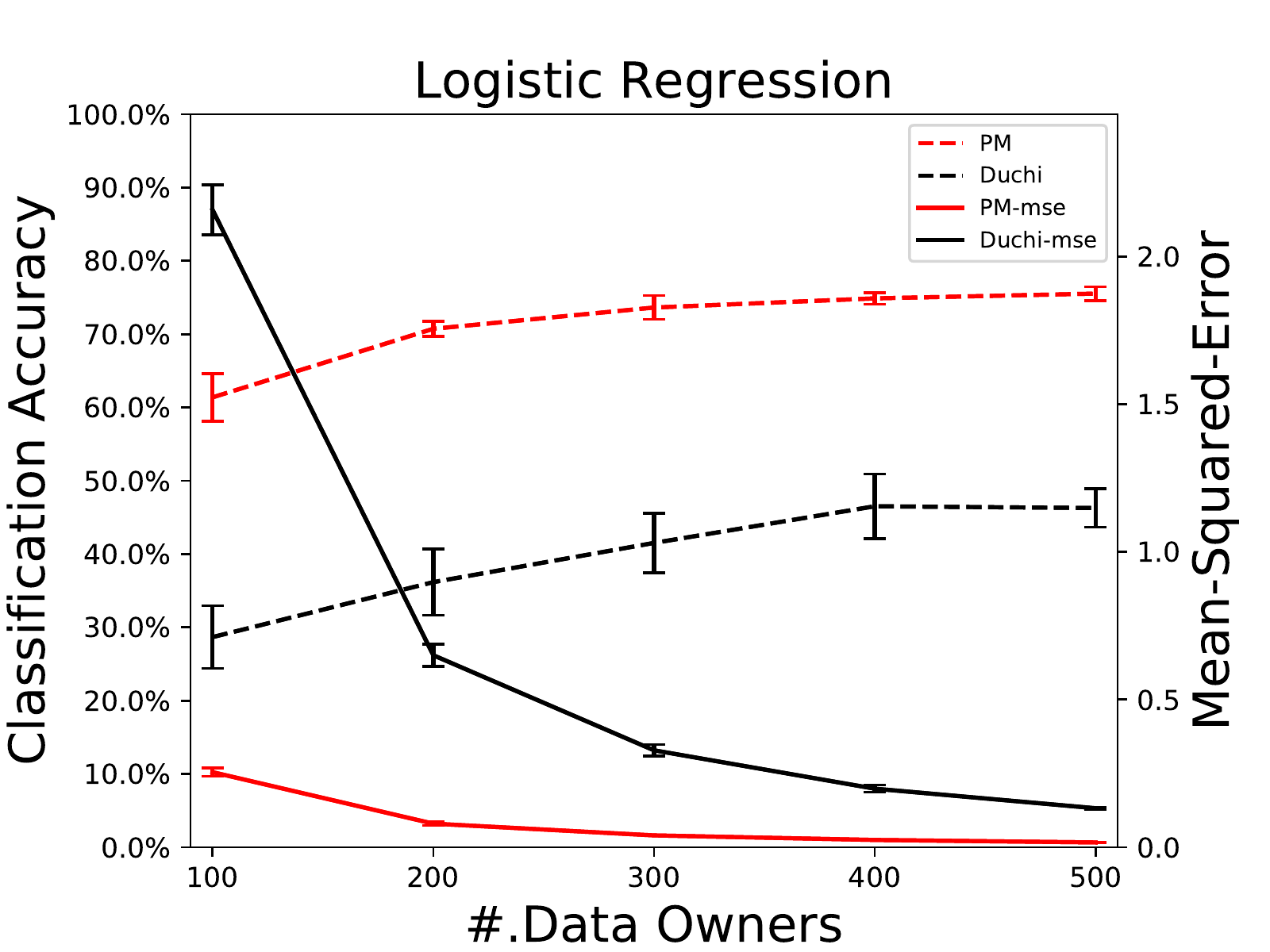}
    \caption{MNIST LR classification accuracy \& MSE v.s. number of data owners, $\epsilon$=5.0, \#.dim = 50}
    \label{fig:mnist_lr_n_dataowner}
\end{minipage}%
    \hfill%
\begin{minipage}[t]{0.45\linewidth}
    \includegraphics[width=\linewidth]{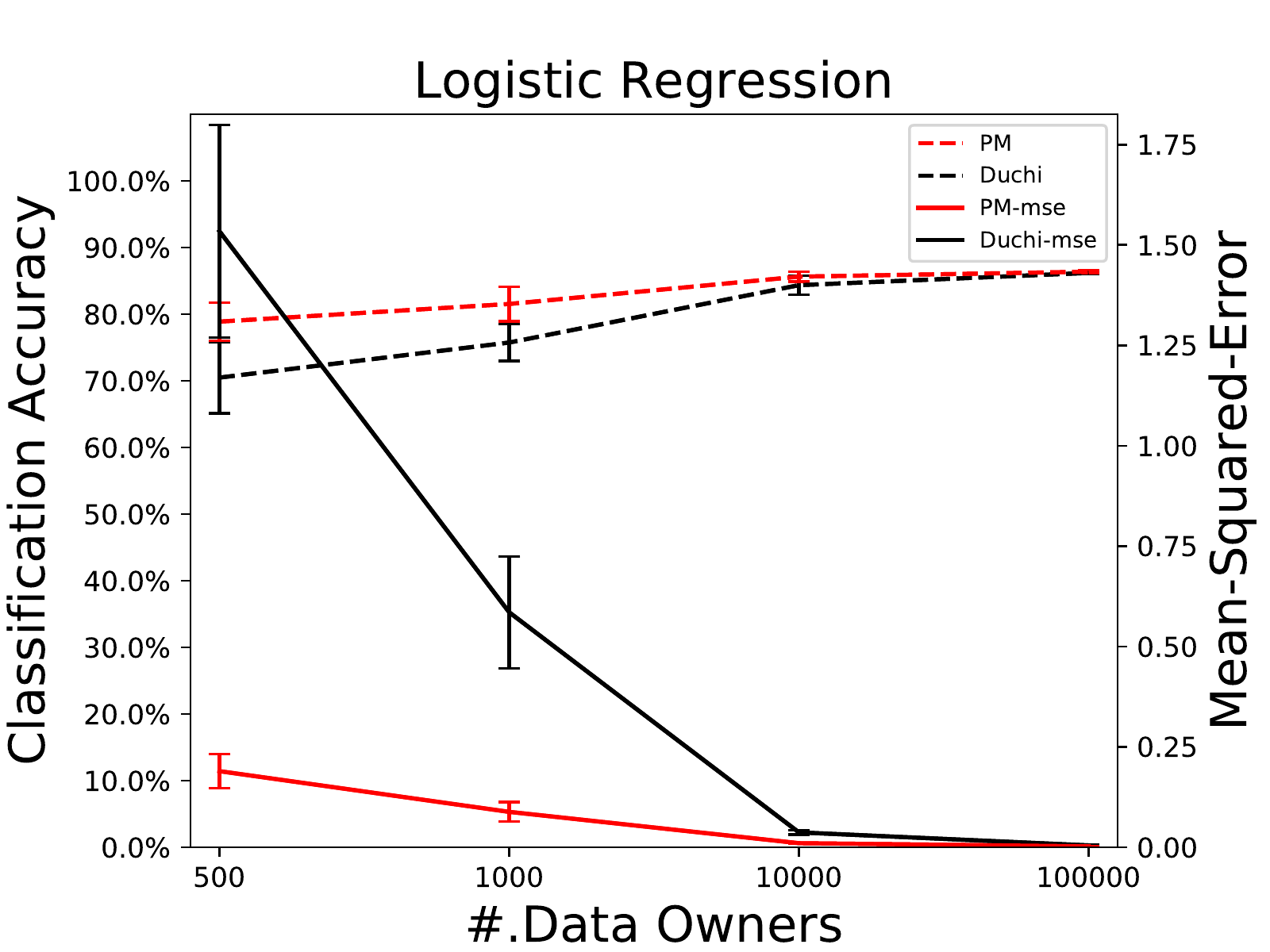}
    \caption{Brazil LR classification accuracy \& MSE v.s. number of data owners, $\epsilon$=5.0, \#.dim = 67}
    \label{fig:brazil_lr_n_dataowner}
\end{minipage}%
\end{figure}

\begin{figure}
  \centering
    \includegraphics[width=0.4\textwidth]{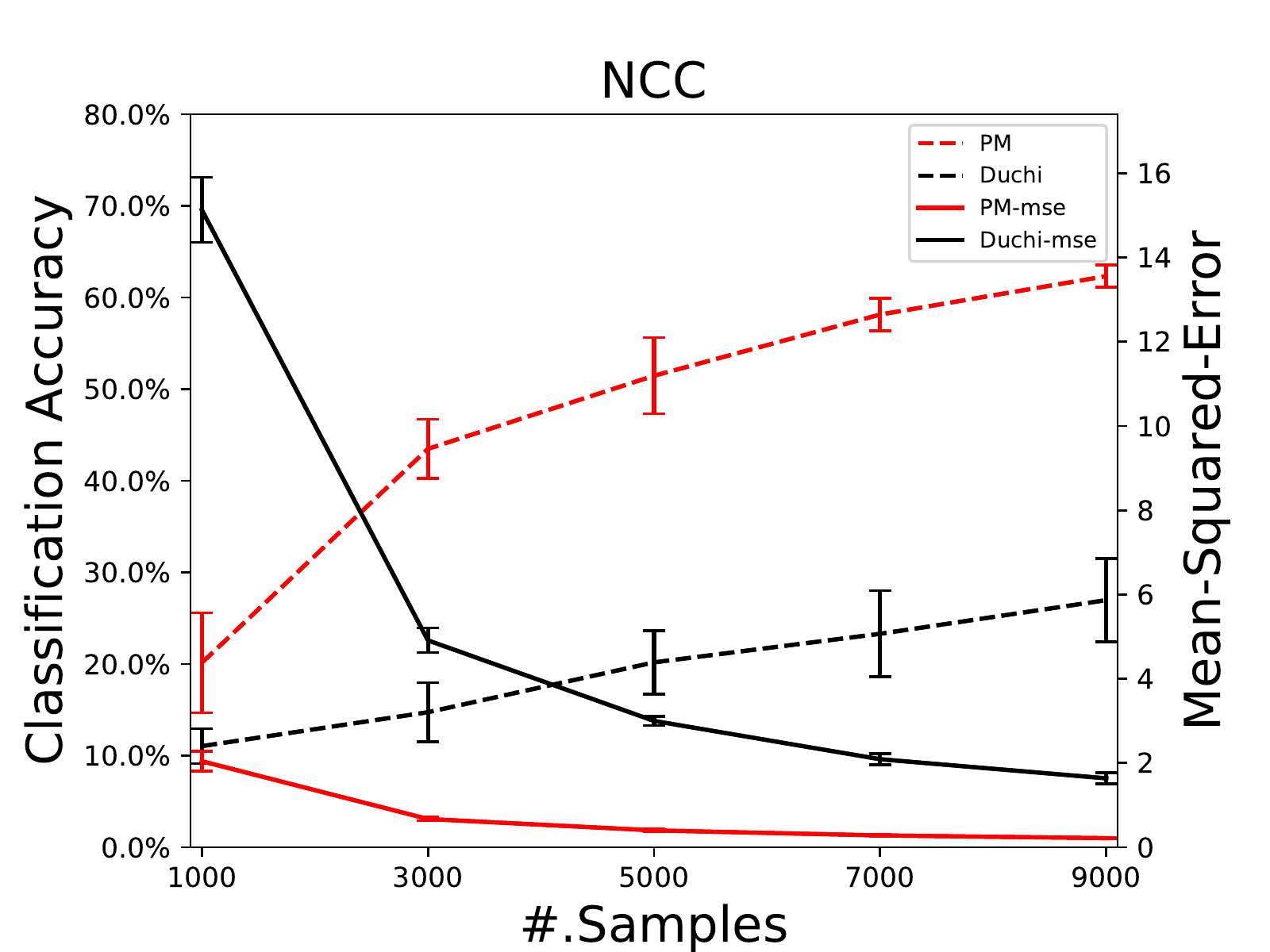}
  \caption{MNIST NCC classification accuracy \& MSE v.s. number of samples, $\epsilon$=3.0, \#.dim=100}
  \label{fig:mnist_ncc_n_samples}
\end{figure}

\begin{figure*}[ht!]
\centering
\begin{subfigure}[t]{1\textwidth}
\begin{subfigure}[t]{.32\textwidth}
\centering
\includegraphics[width=1\textwidth]{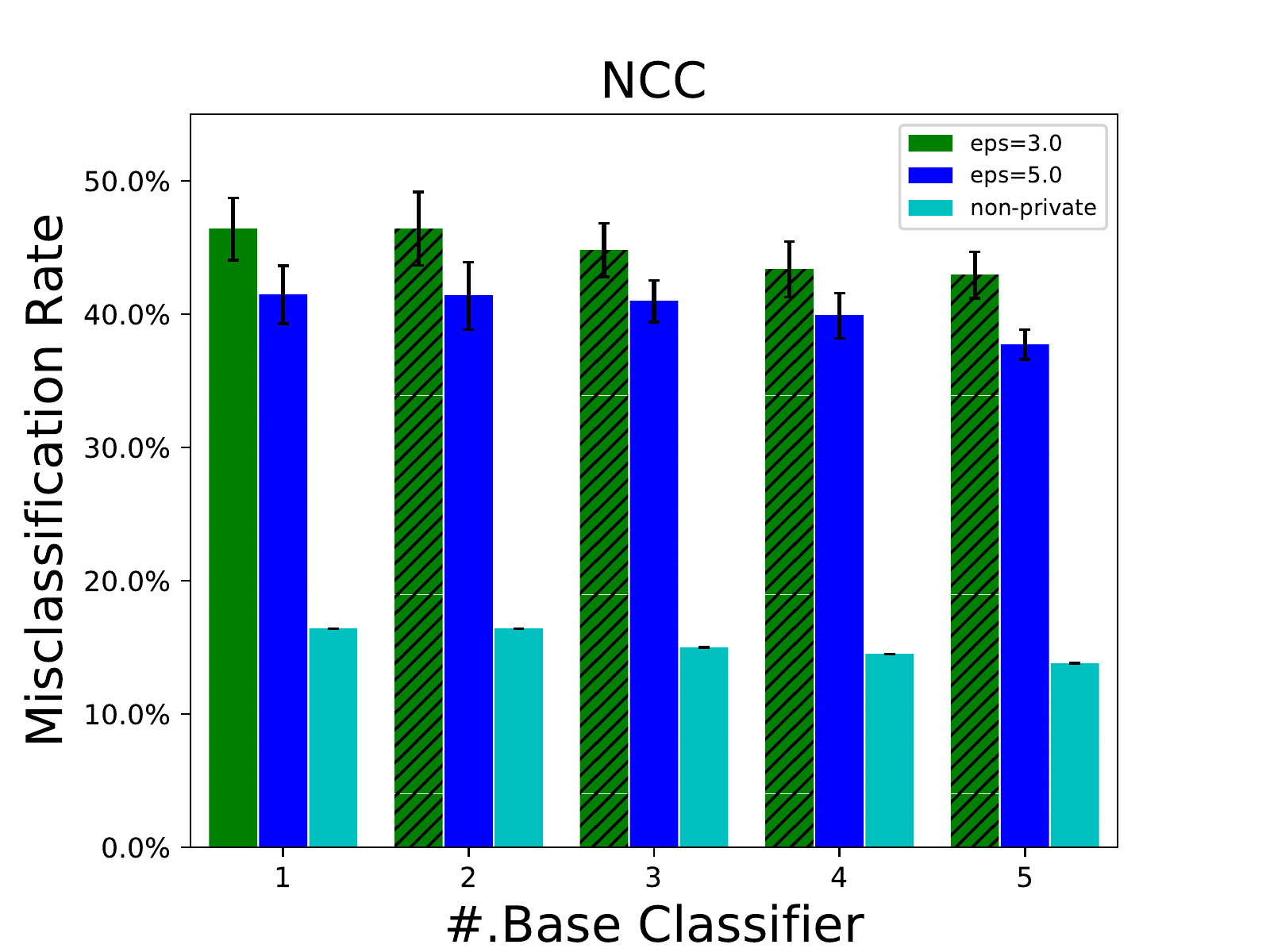}
\caption{MNIST}
\end{subfigure}
\begin{subfigure}[t]{.32\textwidth}
\centering
\includegraphics[width=1\textwidth]{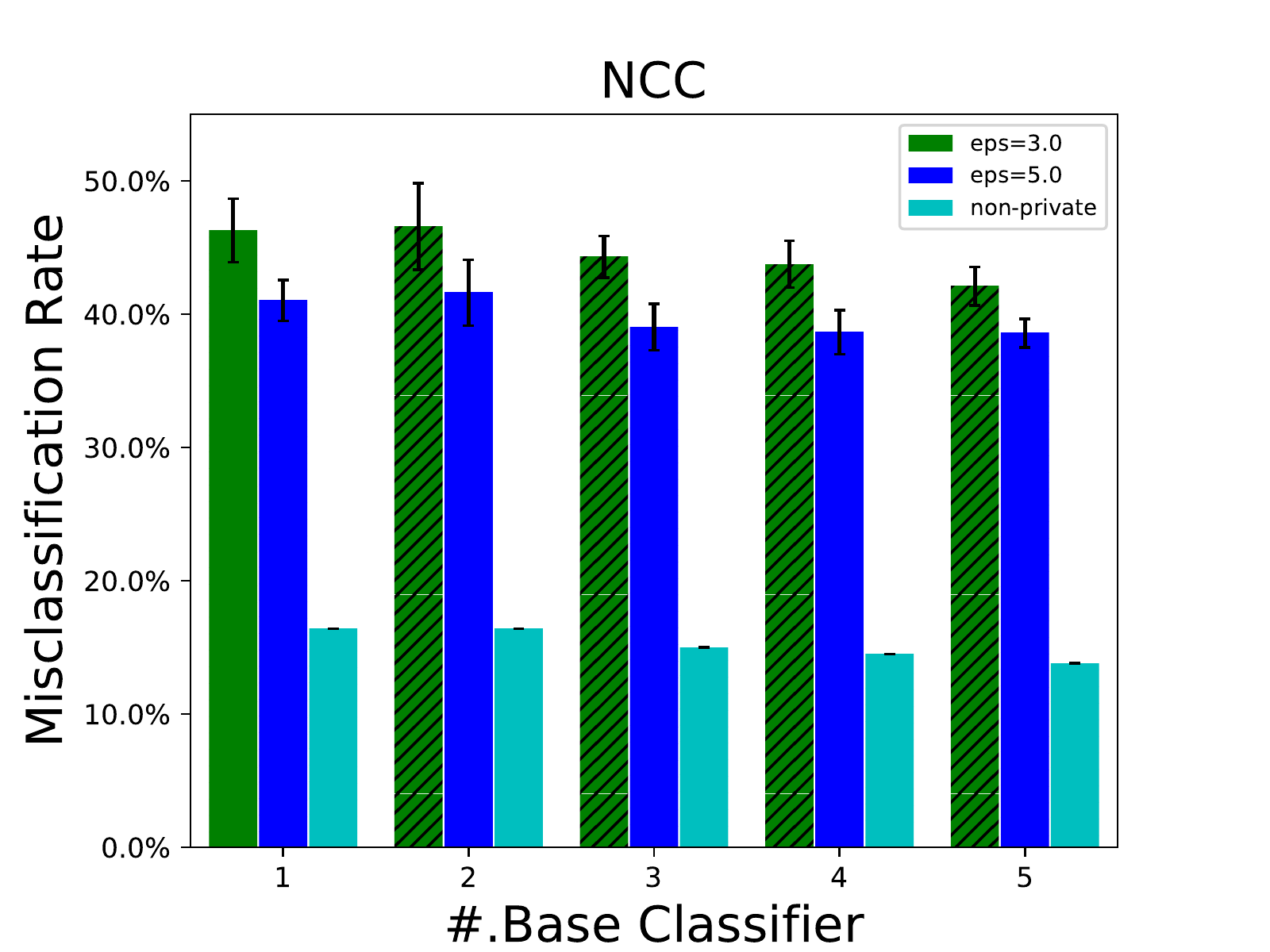}
\caption{Fashion-MNIST}
\end{subfigure}
\begin{subfigure}[t]{.32\textwidth}
\centering
\includegraphics[width=1\textwidth]{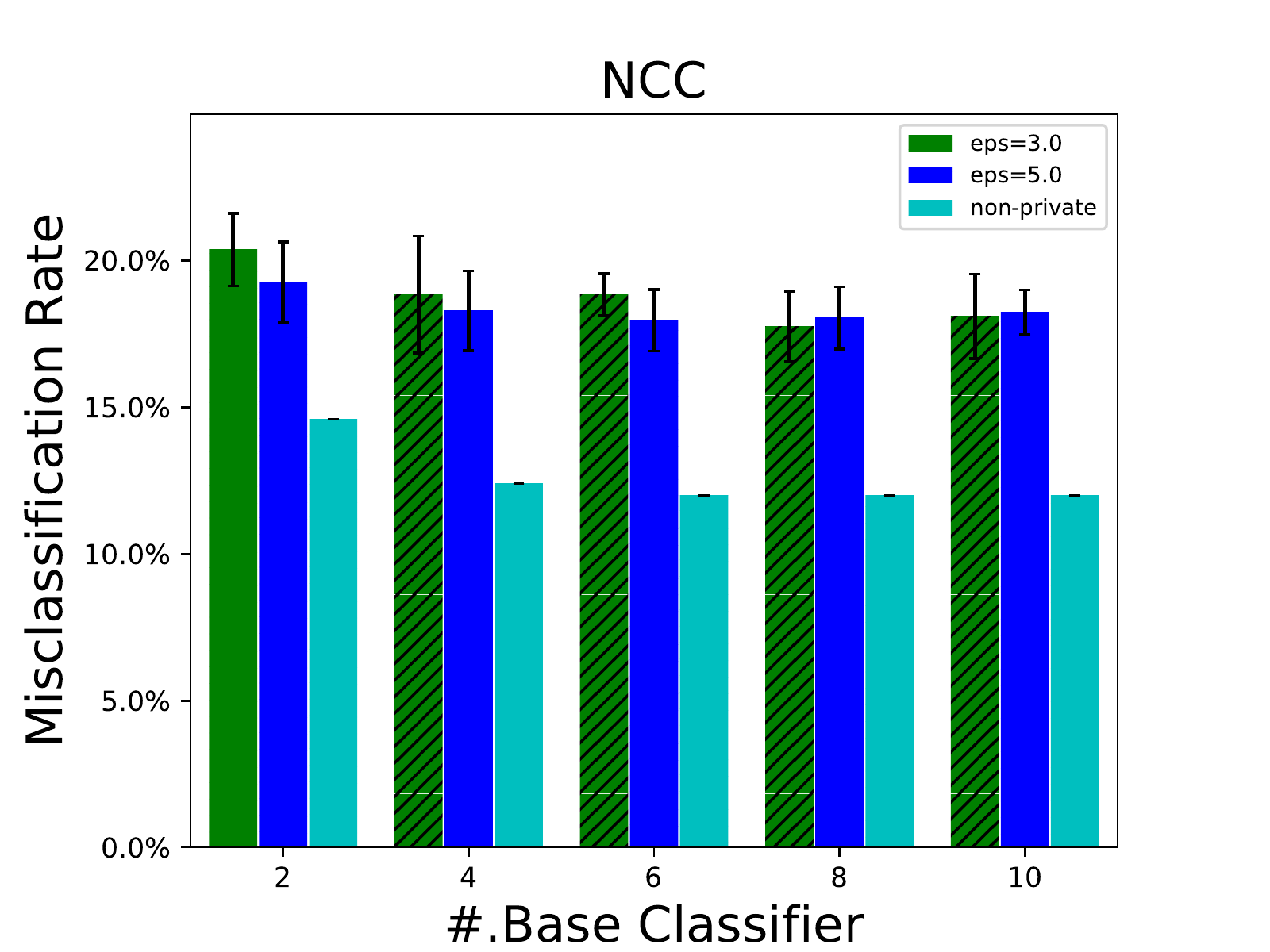}
\caption{Brazil}
\end{subfigure}
\end{subfigure}
\medskip
\caption{Boosted NCC misclassification rate v.s. number of base learners. For MNIST and Fashion-MNIST, the dimensionality of the datasets is 100, and each base learner is fitted by 10,000 perturbed samples. For Brazil, the dimensionality of the dataset is 67, and each base learner is fitted by $10^5$ perturbed samples.}
\label{fig:ncc_n_round}
\end{figure*}

Regarding the number of data owners, recall the total noise incurred in the mean estimation by PM is in $O(\frac{d\sqrt{logd}}{\epsilon\sqrt{n}})$, where $n$ is the number of samples. While aggregating the local classifiers from $L$ data owners, the noise is in $O(\frac{d\sqrt{logd}}{\epsilon\sqrt{L}})$, since each data owner contributes one vector of parameters to represent the local classifier. And the error becomes less as more data owners involve in the aggregation. To illustrate the point, we fix the amount of the privacy budget and varies the number of data owners to see the effect. Fig.~\ref{fig:mnist_lr_n_dataowner} shows the result of the MNIST dataset and Fig.~\ref{fig:brazil_lr_n_dataowner} shows the result of the Brazil dataset. In each figure, the horizontal axis specifies the number of data owners, the left vertical axis shows the classification accuracy and the right vertical axis provides the MSE of the mean of the parameter vectors of the classifiers. From Fig.~\ref{fig:lr}, it concludes that Laplace performs worse than the other two perturbation methods, thus in the figures here we only display the comparison between PM and Duchi. From the MNIST dataset, it can be seen that PM always has less noise than Duchi, and the classifiers of former consistently achieve higher accuracy than the latter. From the Brazil dataset, it can be observed that both PM and Duchi achieves similar loss when the number of data owners grows over 10,000. And the experiment results agree with the theoretical analysis. To summarize, the MSE has a substantial drop from hundreds of data owners to tens of thousands of data owners, while classification accuracy improves with a small magnitude. The reason might be the incurred noise has a small influence to the magnitude of the coefficients of LR, which implies the coefficients of LR has a larger tolerance range to errors.

The experiments above demonstrate the utility of the base classifier in one round of Alg.~\ref{alg:pp_samme}, and we didn't observe accuracy improvements by boosting multiple LR classifiers, the reason might due to the linearity of the model.

\subsection{Nearest Centroid Classifier}
In this subsection, we present the evaluation result of the NCC. Instead of contributing one vector from each data owner, for NCC, each data owner perturbs each individual data sample and submits all perturbed samples he held. The final classifier is learned by finding the centroid of each class after  all perturbed samples have been collected. Fig.~\ref{fig:ncc} shows the experiment result regarding three perturbation methods. And it can be seen that PM performs better than other two perturbation methods for NCC, however, the performance gap between PM and Duchi is larger than the LR, which implies that the parameters of NCC are more sensitive than LR since the decision boundary of NCC is determined by the Euclidean distance between the samples and the centroid, thus the magnitude of the noise has a bigger impact to the prediction capacity. For the Brazil dataset, it can be seen that with the large enough samples sizes (millions), the utility of the learned NCC is well preserved for even with a low privacy budget (1.0). 

Fig.~\ref{fig:mnist_ncc_n_samples} presents the classification result regarding various number of collected samples, ranges from 1000 to 9000. Comparing to LR, the impact of the resulted error in the parameters is more significant, the reason is explained above.

Finally, we boost the NCC classifier with Alg.~\ref{alg:pp_samme} and the result is depicted in Fig.~\ref{fig:ncc_n_round}. Each subfigure shows the result for one dataset, and boosting the NCC leads a $3\%-4\%$ accuracy gain when no noise is injected in general, which proves the effectiveness of the boosting algorithm. Due to the sample size, the number of base classifiers is limit under 10, as each data owner only contribute once during boosting. The figure also plots the accuracy under two privacy budgets, 3.0 and 5.0, and it can be seen that even though the injected noise hurts the utility of the learned base classifier in each round, as the misclassification rate is high compared to the noise-free result, the boosted NCC still gets lower misclassification rate than non-boosted NCC (the first round of boosting), which shows the effectiveness of the proposed algorithm.

\subsection{Boosted Decision Tree}
We present the experiment result of the BDT classifier in this subsection, the evaluation is performed over the MNIST and a synthetic dataset. From the MNIST dataset, we picked the confusable digit 4 and 9 for the classification; For the synthetic dataset, there are two classes and twenty attributes generated, ten of them are the meaningful attributes, the rest of them are generated as non-informative features that are the linear combination of the ten meaningful ones. And there are $10^6$ synthetic samples generated. Both datasets have a balanced number of positive and negative samples. In the experiment, the cross tables of all attributes from each trunk are computed and perturbed, then the cross tables from all trunks are aggregated and the best attribute is decided by the calculated minimum misclassification error. Thus a decision stump is built over the best attribute in each round. Due to the injected noise, the computed attribute from the perturbed cross tables might not be true best attribute. Thus we first investigate the top-k hitting rate of the computed best attribute. Given the non-perturbed cross tables of all attributes, the attributes are sorted based on the misclassification error in an ascendant order. The top-k hitting rate is the probability that the computed best attribute from the perturbed cross tables appears in the top-k non-perturbed best attributes. The larger k is, the higher hitting rate is. When k is equal to the number of the total attributes, the hitting rate reaches the $100\%$. In the experiment, the probability is computed over 100 runs. 

\begin{figure}[htbp]
\begin{minipage}[t]{0.45\linewidth}
    \includegraphics[width=\linewidth]{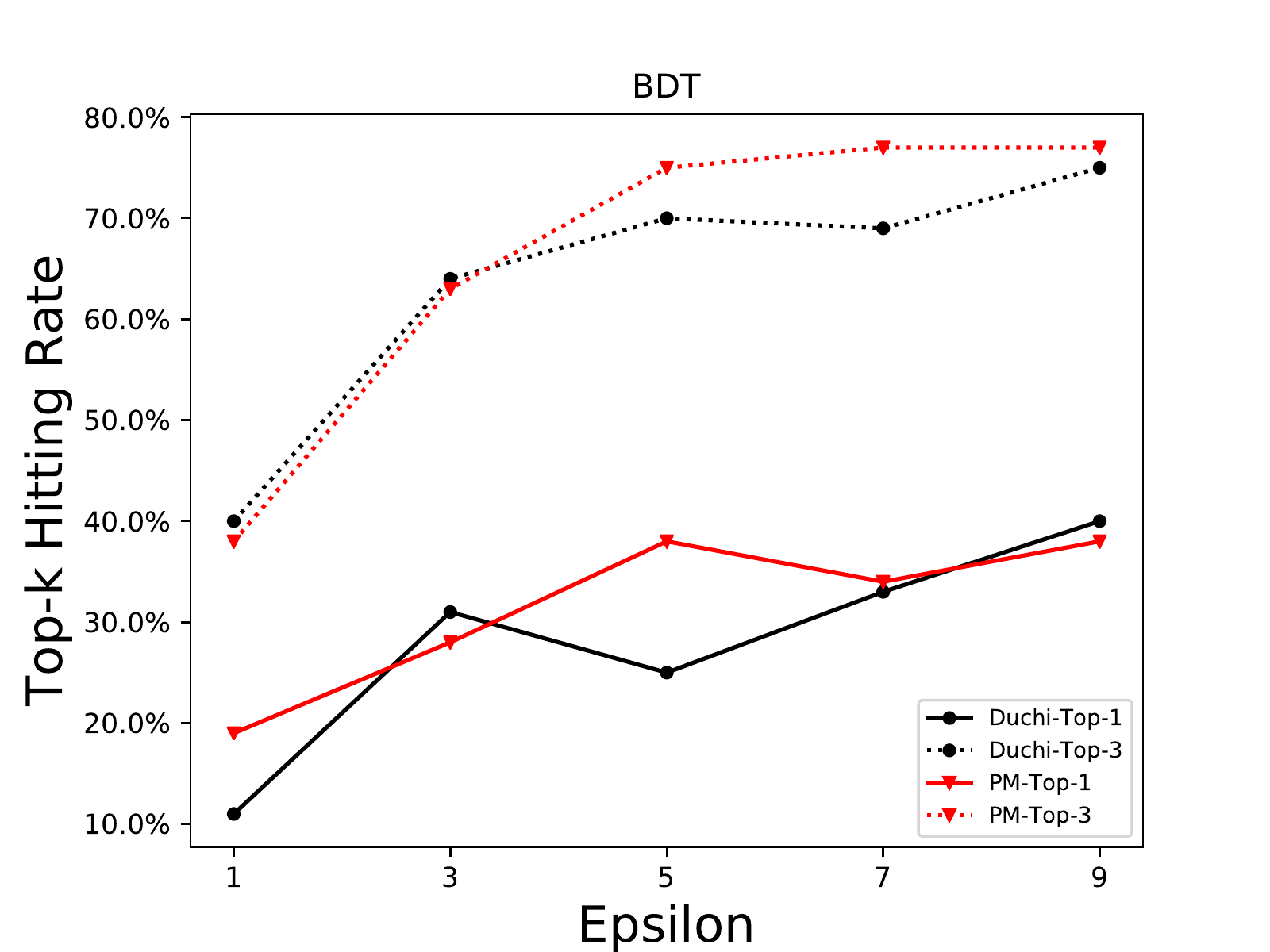}
    \caption{MNIST BDT best attribute top-k hitting rate v.s. privacy budgets}
    \label{fig:mnist_bdt_top_k_hitting}
\end{minipage}%
    \hfill%
\begin{minipage}[t]{0.45\linewidth}
    \includegraphics[width=\linewidth]{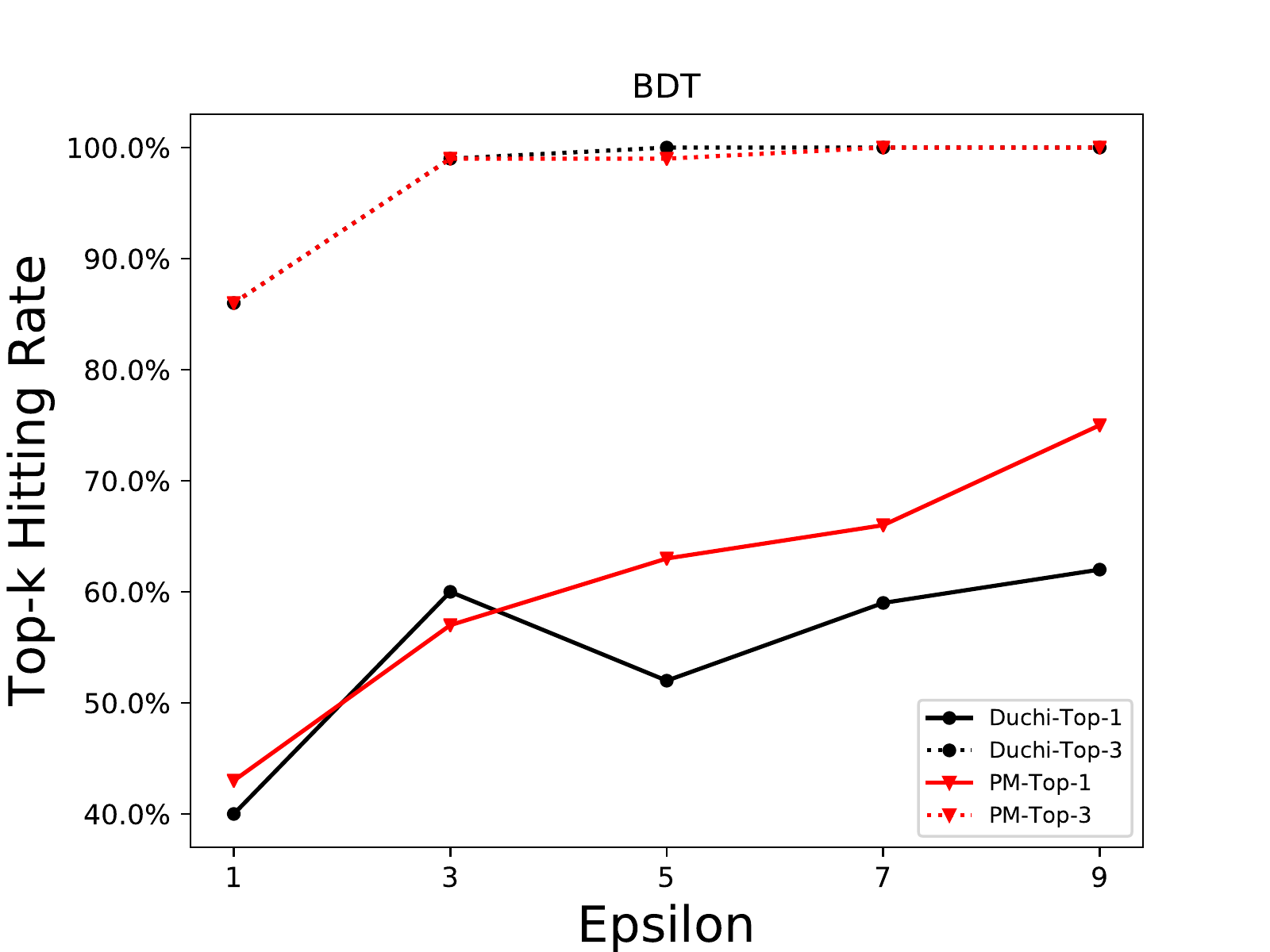}
    \caption{Synthetic dataset BDT best attribute top-k hitting rate v.s. privacy budgets}
    \label{fig:synthetic_bdt_top_k_hitting}
\end{minipage} 
\end{figure}

\begin{figure}[htbp]
\begin{minipage}[t]{0.45\linewidth}
    \includegraphics[width=\linewidth]{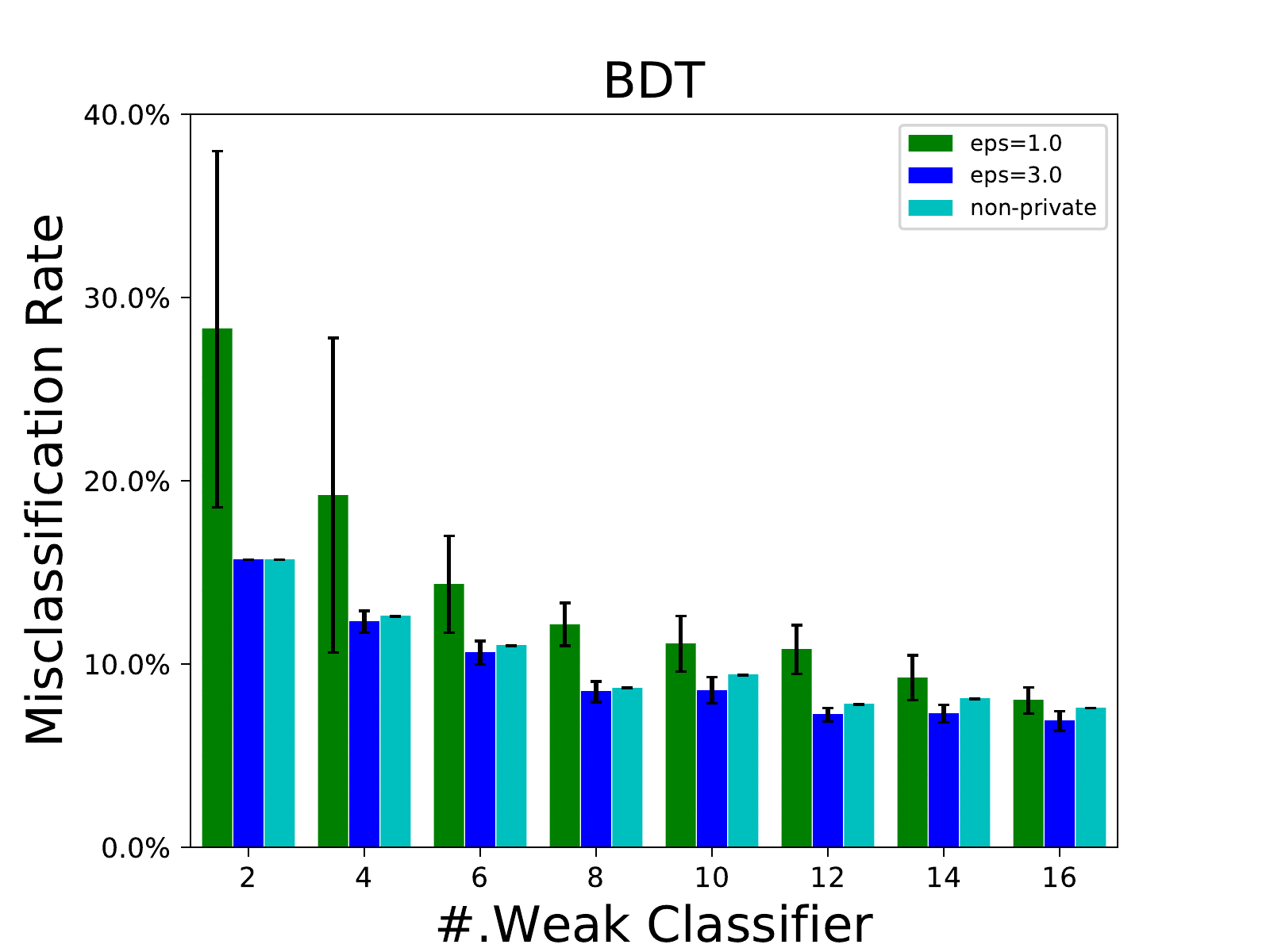}
    \caption{MNIST BDT misclassification rate v.s. number of decision stumps}
    \label{fig:mnist_bdt}
\end{minipage}%
    \hfill%
\begin{minipage}[t]{0.45\linewidth}
    \includegraphics[width=\linewidth]{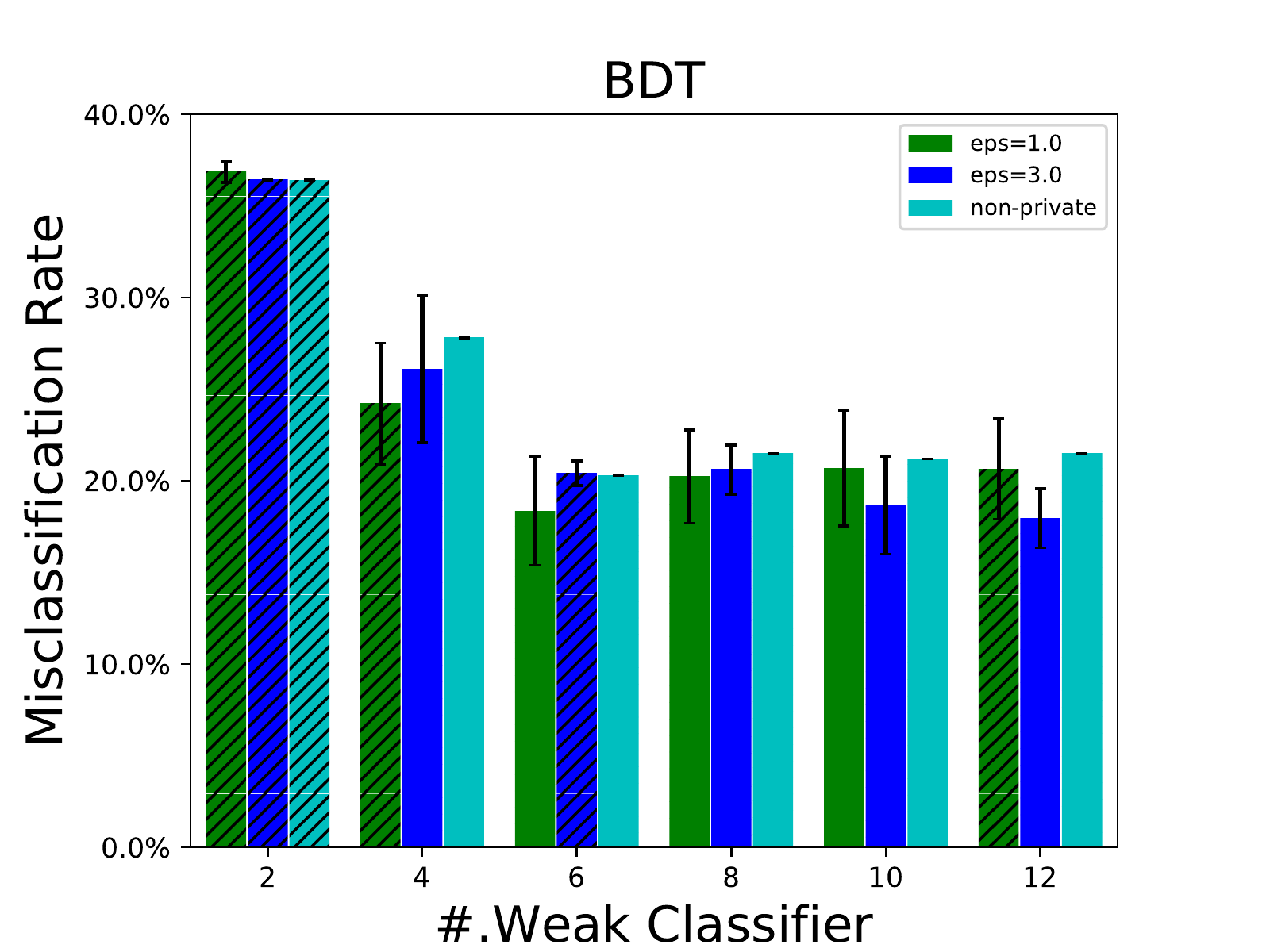}
    \caption{Synthetic dataset BDT misclassification rate v.s. number of decision stumps}
    \label{fig:synthetic_bdt}
\end{minipage} 
\end{figure}

Fig.~\ref{fig:mnist_bdt_top_k_hitting} and Fig.~\ref{fig:synthetic_bdt_top_k_hitting} depict the top-1 and top-3 hitting rates regarding various privacy budgets for the MNIST binary and synthetic dataset. For a comparison, both Duchi and PM perturbation methods are evaluated, and the result are computed for building one decision stump. For both datasets, the top-k hitting rates increase as $\epsilon$ increases, which implies less noise injected and the estimated error become less. For the binary MNIST dataset, the top-3 hitting rates of PM perturbation reaches almost $80\%$ as $\epsilon$ grows to 5.0, which implies that the probability is around $80\%$ that the computed best attributes from the perturbed cross table is one of the top-3 true best attributes; while the top-1 hitting rate is around $40\%$. The reason is that there are 1,000 data owners simulated in the experiment, and the resulted error is larger than the classification error difference among the top-3 best attributes. Similarly, Fig.~\ref{fig:synthetic_bdt_top_k_hitting} shows the result that there are 10,000 data owners simulated from the synthetic dataset, and the top-3 hitting rate is almost $100\%$ as $\epsilon$ reaches to 3.0. For the classification accuracy, Fig.~\ref{fig:mnist_bdt} and ~\ref{fig:synthetic_bdt} plots the misclassification rate as a function of the number of decision stumps in the classifier. For a comparison, the non-private BDT is also depicted in the figure, and it confirms the effectiveness of the boosting algorithm. For the BDT classifier built from the perturbed data, it can be seen that the it has a similar prediction capacity compared to the non-private one under various number of decision stumps. The result confirms that the privacy-preserving BDT successfully maintains a similar utility gain as the non-private one.

\section{Related Work}

Local Differential Privacy (LDP) is firstly proposed by \cite{duchi2013local}, in which it is proposed to eliminate the trusted curator and allow the data owner to control the information releasing in a private manner. Collecting the statistical information is one of great interests in the community, and a randomized response mechanism\cite{warner1965randomized} was invented decades ago to be used in public domain survey. Recently, Wang et al. propose a framework\cite{wang2017locally} that generalizes several LDP protocols in the literature and analyzes each one regarding the task of frequency estimation. Bassily and Smith \cite{bassily2015local} develops an asymptotically optimal solution for the succinct histogram building over the categorical domain under LDP. And for the discrete inputs, Kairouz et al. \cite{kairouz2014extremal} study the trade-off between the utility and LDP in terms of family of extremal mechanisms. Two simple extremal mechanisms, the binary and randomized response mechanisms, are proposed and shown the theoretical guarantee of the utility. For the real-world application, RAPPOR\cite{erlingsson2014rappor} is the first that practically adopts LDP to collect the statistics from the end users, which is designed by Google. Apple and Microsoft also propose their own tools to use LDP in collecting usage, typing history\cite{apple2017dp} and telemetry data\cite{ding2017collecting}. Lately, a \textit{Encode, Shuffle, Analysis} (ESA) system architecture\cite{bittau2017prochlo} is proposed to support a wide range of analysis tasks privately and is used in the software monitoring implementation. All above works focus on the applications based on frequency estimation, where the underlying data are more likely categorical or discrete, while our problem focuses on the numerical data. 

Under LDP, several data mining and machine learning problems have been studied, such as probability distribution estimation\cite{pastore2016locally,kairouz2016discrete,murakami2018toward,ye2018optimal}, frequent itemset mining \cite{wang2018locally}, Bayes learning\cite{yilmaz2019locally} and clustering \cite{nissim2017clustering}. Furthermore, Stochastic Gradient Descent (SGD) that satisfying $\epsilon$-LDP have been used in many applications. Wang et al.\cite{wang2019collect} proposes an piecewise mechanism for multi-dimension numerical attributes perturbation and demonstrate its usage in empirical risk minimization In the paper, the authors demonstrate the computation with three machine learning models, the linear regression, logistic regression and SVM, where multiple data owners locally compute and perturb the gradients of the parameters, then shares the noisy version to the aggregator. Another client-server machine learning framework is proposed to learn the Generalized Linear models\cite{pihur2018differentially}, where the server simultaneous delivers the models to the distributed clients and asks for the parameter update. More specifically, the server maintains $k$ versions of the machine learning model, and randomly selects one version to update the parameters and replaces another version. To enforce the $\epsilon$-LDP, the client uses Laplace mechanism to perturb the parameters and returns the model to the server. Zhang et al. \cite{zhang2017private} consider the problem of multiparty deep learning, wherein multiple data owners train a local deep neural network model and then privately share the model for aggregation, the $\epsilon$-LDP is enforced by the Laplace mechanism before each data owner share the local gradients. The difference between our work and these works is the machine learning model, we are interested in the privacy protection during the boosting procedure and the type of classifiers that could be supported under the privacy-aware manner, for example, the BDT classifier, which is built by aggregating the statistical information and is hardly built through the SGD approach.

Privacy-preserving model sharing and aggregation has been studied in several literature. Pathak et al.\cite{pathak2010multiparty} propose a protocol of composing a differentially private global classifier (e.g., Logistic Regression classifier) by aggregating the classifiers locally trained at the data owner. To protect each individual data sample, the secure protocol utilizes secure multi-party computation (SMC) to aggregate the local classifiers and the global classifier is then perturbed to satisfy the $\epsilon$-DP. Decision tree induction in the distributed setting has been studied in many applications, Caragea et al.\cite{caragea2004framework} formulates the problem of learning from distributed data and proposes a generate strategy to transform the machine learning algorithm into distributed setting. As an example, for the decision tree induction, the local statistics are collected at the centralized site and then the best splitting attribute is decided by maximizing the gain from the aggregated statistics. Another work by Bar-Or et al.\cite{bar2005hierarchical} optimizes the communication overhead of the exchanging statistics by filtering out the attributes based on the bounds of the information gain and develops a distributed hierarchical decision tree algorithm. Bhaduri et al.\cite{bhaduri2008distributed} develops a scalable distribute algorithm for decision tree induction in the peer-to-peer environment, which works in a completely asynchronous manner and has a low communication overhead. Comparing to regular decision tree, Boosted Decision Tree (BDT) limits the depth of the tree to one and shares the same strategy to split the node. Gambs et al.\cite{gambs2007privacy} introduces the privacy-preserving boosting algorithm based on MPC, but the algorithm doesn't consider the information disclosure of individual learning example, thus the technique (e.g., without adding noise for differential privacy) gives inadequate privacy protection, in contrast, our privacy-preserving boosting algorithm provides a stronger solution that prevents the information leakage. 
\section{Conclusion}
In this paper, we propose a privacy-preserving boosting algorithm, and there are two types of parties in our scenario, a data user intends to run the boosting algorithm and learn a classifier and multiple data owners that are willing to contribute the data for the learning purpose but the with privacy protection. The proposed algorithm guarantees that the data user cannot learn the individual training sample from the data owner. The assurance is achieved by the Local Differentially Private (LDP) mechanism, where each data owner perturbs their local data share and then contributes the perturbed data to the data user. Due to the noise injected at individual training sample, only the statistic information of the aggregated data could be well preserved, e.g., mean estimations. With the proposed boosting algorithm, we implement the BDT classifier by utilizing the statistical information from the perturbed data and analyze the privacy and utility trade-off. Furthermore, we also study two other types of classifiers that could be supported and the corresponding data shares to be adopted, e.g., Nearest Centroid Classifier and Logistic Regression. In the experiment, we investigate the utility of the learned classifiers in terms of the classification capacity through multiple real datasets, and compare to two existing perturbation methods, the result shows that the learned classifiers by our method effectively maintain a superior utility.

\bibliographystyle{IEEEtran}
\bibliography{IEEEabrv,./manuscript/citation}

\begin{thebibliography}{10}
\providecommand{\url}[1]{#1}
\csname url@samestyle\endcsname
\providecommand{\newblock}{\relax}
\providecommand{\bibinfo}[2]{#2}
\providecommand{\BIBentrySTDinterwordspacing}{\spaceskip=0pt\relax}
\providecommand{\BIBentryALTinterwordstretchfactor}{4}
\providecommand{\BIBentryALTinterwordspacing}{\spaceskip=\fontdimen2\font plus
\BIBentryALTinterwordstretchfactor\fontdimen3\font minus
  \fontdimen4\font\relax}
\providecommand{\BIBforeignlanguage}[2]{{%
\expandafter\ifx\csname l@#1\endcsname\relax
\typeout{** WARNING: IEEEtran.bst: No hyphenation pattern has been}%
\typeout{** loaded for the language `#1'. Using the pattern for}%
\typeout{** the default language instead.}%
\else
\language=\csname l@#1\endcsname
\fi
#2}}
\providecommand{\BIBdecl}{\relax}
\BIBdecl

\bibitem{leon2012johnny}
P.~Leon, B.~Ur, R.~Shay, Y.~Wang, R.~Balebako, and L.~Cranor, ``Why johnny
  can't opt out: a usability evaluation of tools to limit online behavioral
  advertising,'' in \emph{Proceedings of the SIGCHI Conference on Human Factors
  in Computing Systems}, 2012.

\bibitem{xia2016privacy}
Z.~Xia, X.~Wang, L.~Zhang, Z.~Qin, X.~Sun, and K.~Ren, ``A privacy-preserving
  and copy-deterrence content-based image retrieval scheme in cloud
  computing,'' \emph{IEEE Transactions on Information Forensics and Security},
  2016.

\bibitem{chan2008privacy}
A.~B. Chan, Z.-S.~J. Liang, and N.~Vasconcelos, ``Privacy preserving crowd
  monitoring: Counting people without people models or tracking,'' in
  \emph{IEEE Conference on Computer Vision and Pattern Recognition, CVPR.},
  2008.

\bibitem{lepinski2016privacy}
M.~Lepinski, D.~Levin, D.~McCarthy, R.~Watro, M.~Lack, D.~Hallenbeck, and
  D.~Slater, ``Privacy-enhanced android for smart cities applications,'' in
  \emph{EAI International Conference on Smart Urban Mobility Services}, 2015.

\bibitem{hansen1990neural}
L.~K. Hansen and P.~Salamon, ``Neural network ensembles,'' \emph{IEEE
  Transactions on Pattern Analysis \& Machine Intelligence}, 1990.

\bibitem{breiman1996bagging}
L.~Breiman, ``Bagging predictors,'' \emph{Machine learning}, 1996.

\bibitem{schapire1990strength}
R.~E. Schapire, ``The strength of weak learnability,'' \emph{Machine learning},
  1990.

\bibitem{freund1997decision}
Y.~Freund and R.~E. Schapire, ``A decision-theoretic generalization of on-line
  learning and an application to boosting,'' \emph{Journal of computer and
  system sciences}, 1997.

\bibitem{wolpert1992stacked}
D.~H. Wolpert, ``Stacked generalization,'' \emph{Neural networks}, 1992.

\bibitem{apte1998text}
C.~Apte, F.~Damerau, S.~Weiss \emph{et~al.}, \emph{Text mining with decision
  rules and decision trees}.\hskip 1em plus 0.5em minus 0.4em\relax Citeseer,
  1998.

\bibitem{pal2003assessment}
M.~Pal and P.~M. Mather, ``An assessment of the effectiveness of decision tree
  methods for land cover classification,'' \emph{Remote sensing of
  environment}, 2003.

\bibitem{xia2017boosted}
Y.~Xia, C.~Liu, Y.~Li, and N.~Liu, ``A boosted decision tree approach using
  bayesian hyper-parameter optimization for credit scoring,'' \emph{Expert
  Systems with Applications}, 2017.

\bibitem{shen2007privacy}
X.~Shen, B.~Tan, and C.~Zhai, ``Privacy protection in personalized search,'' in
  \emph{ACM Special Interest Group in Information Retrieval Forum}, 2007.

\bibitem{narayanan2008robust}
A.~Narayanan and V.~Shmatikov, ``Robust de-anonymization of large sparse
  datasets,'' in \emph{IEEE Symposium on Security and Privacy}, 2008.

\bibitem{Dwork2006DP}
C.~Dwork, ``Differential privacy,'' in \emph{33 International Conference on
  Automata, Languages and Programming}, 2006.

\bibitem{ji2014differential}
Z.~Ji, Z.~C. Lipton, and C.~Elkan, ``Differential privacy and machine learning:
  a survey and review,'' \emph{arXiv preprint arXiv:1412.7584}, 2014.

\bibitem{duchi2013local}
J.~C. Duchi, M.~I. Jordan, and M.~J. Wainwright, ``Local privacy and
  statistical minimax rates,'' in \emph{IEEE 54th Annual Symposium on
  Foundations of Computer Science (FOCS)}, 2013.

\bibitem{wang2017locally}
T.~Wang, J.~Blocki, N.~Li, and S.~Jha, ``Locally differentially private
  protocols for frequency estimation,'' in \emph{Proceedings of the 26th USENIX
  Security Symposium}, 2017.

\bibitem{nguyen2016collecting}
T.~T. Nguy{\^e}n, X.~Xiao, Y.~Yang, S.~C. Hui, H.~Shin, and J.~Shin,
  ``Collecting and analyzing data from smart device users with local
  differential privacy,'' \emph{arXiv preprint arXiv:1606.05053}, 2016.

\bibitem{pihur2018differentially}
V.~Pihur, A.~Korolova, F.~Liu, S.~Sankuratripati, M.~Yung, D.~Huang, and
  R.~Zeng, ``Differentially-private 'draw and discard' machine learning,''
  \emph{arXiv preprint arXiv:1807.04369}, 2018.

\bibitem{warner1965randomized}
S.~L. Warner, ``Randomized response: A survey technique for eliminating evasive
  answer bias,'' \emph{Journal of the American Statistical Association}, 1965.

\bibitem{kairouz2014extremal}
P.~Kairouz, S.~Oh, and P.~Viswanath, ``Extremal mechanisms for local
  differential privacy,'' in \emph{Advances in neural information processing
  systems}, 2014.

\bibitem{dwork2014analyze}
C.~Dwork, K.~Talwar, A.~Thakurta, and L.~Zhang, ``Analyze gauss: optimal bounds
  for privacy-preserving principal component analysis,'' in \emph{46th Annual
  ACM Symposium on Theory of Computing}, 2014.

\bibitem{hastie2009multi}
T.~Hastie, S.~Rosset, J.~Zhu, and H.~Zou, ``Multi-class adaboost,''
  \emph{Statistics and its Interface}, 2009.

\bibitem{kantarcioglu2004privacy}
M.~Kantarcioglu and C.~Clifton, ``Privacy-preserving distributed mining of
  association rules on horizontally partitioned data,'' \emph{IEEE Transactions
  on Knowledge \& Data Engineering}, 2004.

\bibitem{shamir1979share}
A.~Shamir, ``How to share a secret,'' \emph{Communications of the ACM}, 1979.

\bibitem{pathak2010multiparty}
M.~Pathak, S.~Rane, and B.~Raj, ``Multiparty differential privacy via
  aggregation of locally trained classifiers,'' in \emph{Advances in Neural
  Information Processing Systems}, 2010.

\bibitem{wang2019collect}
N.~Wang, X.~Xiao, Y.~Yang, J.~Zhao, S.~C. Hui, H.~Shin, J.~Shin, and G.~Yu,
  ``Collecting and analyzing multidimensional data with local differential
  privacy,'' in \emph{IEEE 35th International Conference on Data Engineering
  (ICDE)}, 2019.

\bibitem{duchi2018minimax}
J.~C. Duchi, M.~I. Jordan, and M.~J. Wainwright, ``Minimax optimal procedures
  for locally private estimation,'' \emph{Journal of the American Statistical
  Association}, 2018.

\bibitem{mcsherry2007mechanism}
F.~McSherry and K.~Talwar, ``Mechanism design via differential privacy,'' in
  \emph{FOCS}, 2007.

\bibitem{hamm2015crowd}
J.~Hamm, A.~C. Champion, G.~Chen, M.~Belkin, and D.~Xuan, ``Crowd-ml: A
  privacy-preserving learning framework for a crowd of smart devices,'' in
  \emph{IEEE 35th International Conference on Distributed Computing Systems},
  2015.

\bibitem{quinlan1986induction}
J.~R. Quinlan, ``Induction of decision trees,'' \emph{Machine learning}, 1986.

\bibitem{breiman2017classification}
L.~Breiman, \emph{Classification and regression trees}.\hskip 1em plus 0.5em
  minus 0.4em\relax Routledge, 2017.

\bibitem{tan2018introduction}
P.-N. Tan, \emph{Introduction to data mining}.\hskip 1em plus 0.5em minus
  0.4em\relax Pearson Education India, 2018.

\bibitem{bhaduri2008distributed}
K.~Bhaduri, R.~Wolff, C.~Giannella, and H.~Kargupta, ``Distributed
  decision-tree induction in peer-to-peer systems,'' \emph{Statistical Analysis
  and Data Mining: The ASA Data Science Journal}, 2008.

\bibitem{ipums}
``Integrated public use microdata series,'' \url{https://www.ipums.org}, 2019.

\bibitem{dwork2006calibrating}
C.~Dwork, F.~McSherry, K.~Nissim, and A.~Smith, ``Calibrating noise to
  sensitivity in private data analysis,'' in \emph{Theory of cryptography
  conference}.\hskip 1em plus 0.5em minus 0.4em\relax Springer, 2006.

\bibitem{bassily2015local}
R.~Bassily and A.~Smith, ``Local, private, efficient protocols for succinct
  histograms,'' in \emph{Proceedings of the 47 annual ACM symposium on Theory
  of computing}, 2015.

\bibitem{erlingsson2014rappor}
{\'U}.~Erlingsson, V.~Pihur, and A.~Korolova, ``Rappor: Randomized aggregatable
  privacy-preserving ordinal response,'' in \emph{Proceedings of the 2014 ACM
  SIGSAC conference on computer and communications security}, 2014.

\bibitem{apple2017dp}
D.~P. Team, ``Learning with privacy at scale,'' 2017.

\bibitem{ding2017collecting}
B.~Ding, J.~Kulkarni, and S.~Yekhanin, ``Collecting telemetry data privately,''
  in \emph{Advances in Neural Information Processing Systems}, 2017.

\bibitem{bittau2017prochlo}
A.~Bittau, {\'U}.~Erlingsson, P.~Maniatis, I.~Mironov, A.~Raghunathan, D.~Lie,
  M.~Rudominer, U.~Kode, J.~Tinnes, and B.~Seefeld, ``Prochlo: Strong privacy
  for analytics in the crowd,'' in \emph{Proceedings of the 26th Symposium on
  Operating Systems Principles}.\hskip 1em plus 0.5em minus 0.4em\relax ACM,
  2017.

\bibitem{pastore2016locally}
A.~Pastore and M.~Gastpar, ``Locally differentially-private distribution
  estimation,'' in \emph{IEEE International Symposium on Information Theory
  (ISIT)}, 2016.

\bibitem{kairouz2016discrete}
P.~Kairouz, K.~Bonawitz, and D.~Ramage, ``Discrete distribution estimation
  under local privacy,'' \emph{Proceedings of the 33rd International Conference
  on Machine Learning}, 2016.

\bibitem{murakami2018toward}
T.~Murakami, H.~Hino, and J.~Sakuma, ``Toward distribution estimation under
  local differential privacy with small samples,'' \emph{Proceedings on Privacy
  Enhancing Technologies}, 2018.

\bibitem{ye2018optimal}
M.~Ye and A.~Barg, ``Optimal schemes for discrete distribution estimation under
  locally differential privacy,'' \emph{IEEE Transactions on Information
  Theory}, 2018.

\bibitem{wang2018locally}
T.~Wang, N.~Li, and S.~Jha, ``Locally differentially private frequent itemset
  mining,'' in \emph{IEEE Symposium on Security and Privacy (SP)}, 2018.

\bibitem{yilmaz2019locally}
E.~Yilmaz, M.~Al-Rubaie, and J.~M. Chang, ``Locally differentially private
  naive bayes classification,'' \emph{arXiv preprint arXiv:1905.01039}, 2019.

\bibitem{nissim2017clustering}
K.~Nissim and U.~Stemmer, ``Clustering algorithms for the centralized and local
  models,'' \emph{arXiv preprint arXiv:1707.04766}, 2017.

\bibitem{zhang2017private}
X.~Zhang, S.~Ji, H.~Wang, and T.~Wang, ``Private, yet practical, multiparty
  deep learning,'' in \emph{IEEE 37th International Conference on Distributed
  Computing Systems (ICDCS)}, 2017.

\bibitem{caragea2004framework}
D.~Caragea, A.~Silvescu, and V.~Honavar, ``A framework for learning from
  distributed data using sufficient statistics and its application to learning
  decision trees,'' \emph{International Journal of Hybrid Intelligent Systems},
  2004.

\bibitem{bar2005hierarchical}
A.~Bar-Or, D.~Keren, A.~Schuster, and R.~Wolff, ``Hierarchical decision tree
  induction in distributed genomic databases,'' \emph{IEEE Transactions on
  Knowledge and Data Engineering}, 2005.

\bibitem{gambs2007privacy}
S.~Gambs, B.~K{\'e}gl, and E.~A{\"\i}meur, ``Privacy-preserving boosting,''
  \emph{Data Mining and Knowledge Discovery}, 2007.

\end{thebibliography}


\begin{thebibliography}{1}

\bibitem{IEEEhowto:kopka}
H.~Kopka and P.~W. Daly, \emph{A Guide to \LaTeX}, 3rd~ed.\hskip 1em plus
  0.5em minus 0.4em\relax Harlow, England: Addison-Wesley, 1999.

\end{thebibliography}
%
%

\begin{IEEEbiography}[{\includegraphics[width=1in,height=1.25in,clip,keepaspectratio]{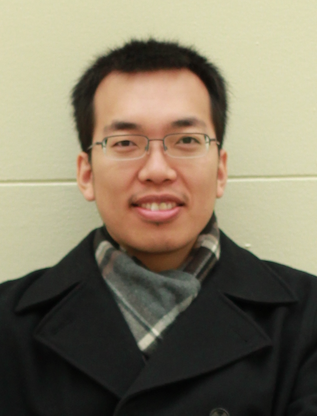}}]{Sen Wang}
received the B.S. degree in software engineering from Xiamen University, China, in 2010, and the M.S. degree in computer science from Iowa State University, in 2013. He is currently pursuing the Ph.D. degree in the Department of Electrical Engineering, University of South Florida, Tampa, Florida. His research interests include the database, security, privacy-preserving data mining and machine learning.
\end{IEEEbiography}
\begin{IEEEbiography}[{\includegraphics[width=1in,height=1.25in,clip,keepaspectratio]{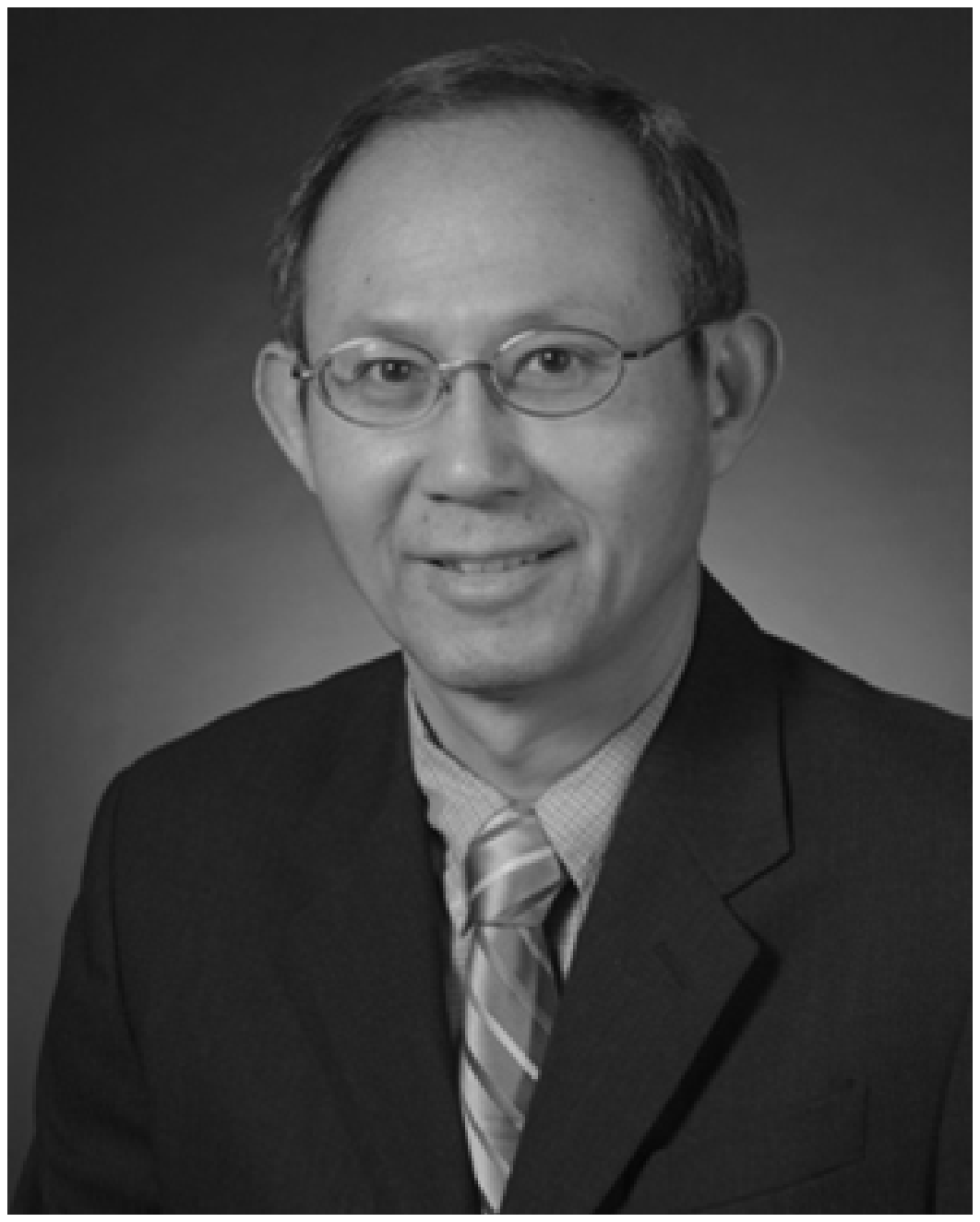}}]{J.Morris Chang} received his BSEE degree from Tatung Institute of Technology, Taiwan and his MS and PhD in Computer Engineering is from North Carolina State University. He is currently a Professor at the Department of Electrical Engineering at University of South Florida. Dr.Changs industrial experience includes positions at Texas Instruments, Taiwan, Microelectronics Center of North Carolina, and AT\&T Bell Laboratories, Pennsylvania. He was on the faculty of the Department of Electrical Engineering at Rochester Institute of Technology, Rochester, the Department of Computer Science at Illinois Institute of Technology, Chicago and the Department of Electrical and Computer Engineering at Iowa State University, IA. His research interests include cyber security, wireless networks, energy-aware computing and object-oriented systems. Currently, Dr. Chang is a handling editor of Journal of Microprocessors and Microsystems and the Associate Editor-in-Chief of IEEE IT Professional. He is a senior member of IEEE.
\end{IEEEbiography}

\end{document}